\documentclass[openacc]{rsproca_new}

\usepackage[normalem]{ulem}
\usepackage[usenames, dvipsnames]{xcolor}

\providecommand*{\deriv}[3][]{%
\dfrac{\mathrm{d}^{#1}#2}{\mathrm{d} #3^{#1}}}
\providecommand*{\pderiv}[3][]{%
\dfrac{\partial^{#1}#2}%
{\partial {#3}^{#1}}}
\providecommand*{\nuchar}{%
\dfrac{\nu}{\nu+\nu_{TC}}}
\providecommand*{\nuTchar}{%
\dfrac{\nu_{TC}}{\nu+\nu_{TC}}}

\newcommand*{\Eref}[1]{(\ref{#1})}
\newcommand*{\Cref}[1]{Equation~(\ref{#1})}

\newcommand*{\revone}[1]{{{\color{black}{#1}}}}
\newcommand*{\revtwo}[1]{{{\color{black}{#1}}}}
\newcommand*{\rev}[1]{{{\color{black}{#1}}}}

\begin{document}

\title{Prandtl's extended mixing length model applied to the two-dimensional turbulent classical far wake}

\author{
 A. J. Hutchinson$^{1}$, N. Hale$^{2}$, K. Born$^{3}$, and D. P. Mason$^{4}$ }

\address{$^{1,3,4}$School of Computer Science and Applied Mathematics, University of the  Witwatersrand, Johannesburg, Private Bag 3, Wits 2050, South Africa.

$^{1,4}$DSI-NRF Centre of Excellence in Mathematical and Statistical Sciences,  South Africa.

$^{2}$Department of Mathematical Sciences, Stellenbosch University, Stellenbosch, 7602, South Africa.}

\subject{Applied Mathematics, Mechanics}


\corres{A. J. Hutchinson

\email{Ashleigh.Hutchinson@wits.ac.za}}

\begin{abstract}
Despite its limitations, Prandtl's mixing length model is widely applied in modelling turbulent free shear flows. Prandtl's extended model addresses many of the shortfalls of the original model, but is not so widely used, in part due to additional mathematical complexities that arise in its derivation and implementation. Furthermore, in both models Prandtl neglects the kinematic viscosity on the basis that it is much smaller in magnitude than the turbulent viscosity. Recent work has shown that including the kinematic viscosity in the original model has both mathematical and physical advantages. In the present work, a \revtwo{novel} derivation of the extended model is provided, and it is demonstrated that similar advantages are again obtained when the kinematic viscosity is included.
Additionally, through the use of scaling techniques, similarity mean velocity profiles of the extended model are derived, resulting in a single nonlinear ordinary differential equation that is solved numerically with a Hermite spectral method. The computed profiles for the normalised similarity mean velocity \revone{and shear stress} are compared to experimental observations and shown to be in excellent agreement.
\end{abstract}


\maketitle

\section{Introduction}
Turbulent flows are ubiquitous in both nature and industry~\cite{Ten,P}.
In many applications, reliable models are needed to predict the behaviour of complex free shear flows. For example, understanding the development of turbulent wakes behind wind turbines allows informed decision-making regarding turbine placement and wake steering control operations~\cite{Turbine}.
Although the use of numerical techniques to simulate turbulence, such as large eddy simulations (LES) and direct numerical simulations (DNS)~\cite{Breton, Mehta,winda,windb,windc,windd}, is increasing, simple analytic models, such as those investigated in the present work, continue to be widely used in the context of free shear flows~\cite{Hutter2016, Cafiero, Note}.

A common approach to modelling turbulence is to derive equations for the mean flow variables \cite{Reynolds,Liepmann}. The Reynolds decomposition, in which a turbulent flow is represented as the combination of a mean flow and a fluctuation, is substituted into the Navier--Stokes equations and the time average is taken. This procedure gives rise to the Reynolds-averaged Navier--Stokes (RANS) equations, which contain unknown turbulent stress terms known as the Reynolds (or {apparent}) stresses~\cite{Reynolds}.
To calculate the mean flow variables from the RANS equations, a closure model is needed. 
Boussinesq proposed the eddy viscosity approach, which relates the turbulent stresses to the mean rates of deformation  \cite[pp.~23-46]{Boussinesq1}.

In this work, particular attention is paid to simple analytic models using the eddy viscosity approach applied to the two-dimensional turbulent classical far wake. Classical far wake studies for laminar flows date back to the 1930's and are largely accredited to Goldstein \cite{Goldstein1933}. A classical wake develops in the region downstream of a stationary solid body placed in a laminar mainstream flow. For sufficiently large Reynolds numbers, these laminar flows become turbulent~\cite{Liepmann}. In algebraic closure models, the effective viscosity is defined as the sum of the kinematic viscosity, which is an intrinsic property of the fluid, and a turbulent or eddy viscosity, which is a characteristic of the flow. Various closure models may then be used to describe the eddy viscosity. The simplest closure model, where the eddy viscosity is taken to be constant, is often used as a baseline to compare against the other models \cite{P, Cafiero, Ten}. However, upon comparison with experimental results, the constant eddy viscosity (CEV) model fails to capture the correct behaviour near the  boundaries of the wake~\cite{Wygnanski}.

To improve upon the CEV model, Prandtl introduced the concept of a mixing length~\cite{Prandtl}, and Prandtl's mixing length (PML) closure model has since been used extensively to describe the eddy viscosity. In this model, the eddy viscosity is written in terms of a mixing length and the gradient of the mean velocity deficit perpendicular to the axis of the wake.  Although an improvement on the CEV model, the PML model still fails to capture some of the important physics observed in experimental data~\cite{Swain} and has the nonphysical property of the eddy viscosity vanishing on the centre line of the wake~\cite{Prandtl}.
Prandtl realised the limitations of his closure model and proposed a modification, which we refer to as the extended Prandtl mixing length (EPML) model, to address some of them. Here, two mixing lengths are introduced and the eddy viscosity is expressed as a function of these mixing lengths and both the gradient and curvature of the mean velocity deficit perpendicular to the centre line of the wake~\cite{Prandtl}. As a result, the eddy viscosity no longer vanishes on the axis of the wake and one of the limitations of the PML model is resolved. However, as with the PML model, it is not possible to obtain the form of the mixing lengths without imposing an additional hypothesis; for example, that the mixing length is proportional to the width of the wake~\cite{PrandtlHypothesis}. Furthermore, although simpler versions of the EPML model have received some attention~\cite{Luo}, {and an initial study pertaining to wake flows has been undertaken~\cite{ashex}}, the model is seldom used in full due to the additional mathematical complexity that results from its implementation \cite{Note}. 

Many of the mathematical and physical limitations of the PML and EPML models arise from one important assumption: the kinematic viscosity can be neglected in comparison to the turbulent viscosity. Mathematically, this assumption does not hold on the axis of the wake for the PML model nor at the wake boundaries for both the PML and EPML models, and as a result, the width of the wake is underestimated \cite{Prandtl}. 

Prandtl's original model was recently modified to include the kinematic viscosity~\cite{ash3}. It was shown that the mixing length can be derived without imposing any additional hypotheses and that the wake boundary predicted from this model lies outside of the underestimated boundary obtained from the original PML model where the kinematic viscosity is neglected. In the present work, we study the effect of including the kinematic viscosity in the EPML model. In particular, we provide a detailed derivation of the EPML model and show that if (and only if) the kinematic viscosity is included then a form for both of the mixing lengths can be obtained without imposing any additional assumptions.  
Furthermore, the derivation is unified in the sense that the PML model appears as a special case, allowing for a convenient comparison. Similarity solutions are used to reduce the governing partial differential equation (PDE) to a second order nonlinear ordinary differential equation (ODE), which can be solved analytically in one case (PML with no kinematic viscosity) and numerically in others (PML with kinematic viscosity, and EPML). The resulting self-similar mean velocity \revone{and shear stress} profiles are compared with experimental data from the literature, and the EPML model is shown to give excellent correspondence.

\revone{The practical use of Lie groups in turbulence modelling has been previously demonstrated, such as the similarity transforms derived by Cantwell~\cite{cantwell_1978} for the two-dimensional unsteady, stream function equation, and the many notable studies showing significant progress in symmetry methods applied to turbulent flows \cite{oberlack_2001,oberlack_2007,oberlack_2015,Ibragimov,Unal}. The similarity methods employed in the current work present yet another example and establish the potential for the general application of Lie Group Theory in evaluating turbulence models.}

The outline of this paper is as follows. In Section \ref{sec2} we present a mathematical model for a two-dimensional turbulent classical far wake. In Section \ref{sec:closure} we present a new derivation of the EPML model with kinematic viscosity included, and in Section \ref{sec4} identify a scaling solution that reduces the model to a one-dimensional ODE. In Section \ref{sec5} we compare the numerically computed solution of the EPML model to that of various other closure models and with experimental results from the literature. A short summary is presented in Section \ref{sec6}.



\section{Mathematical model for a two-dimensional turbulent classical far wake}\label{sec2}

\begin{figure}[ht]
    \centering
    \includegraphics[scale=0.7]{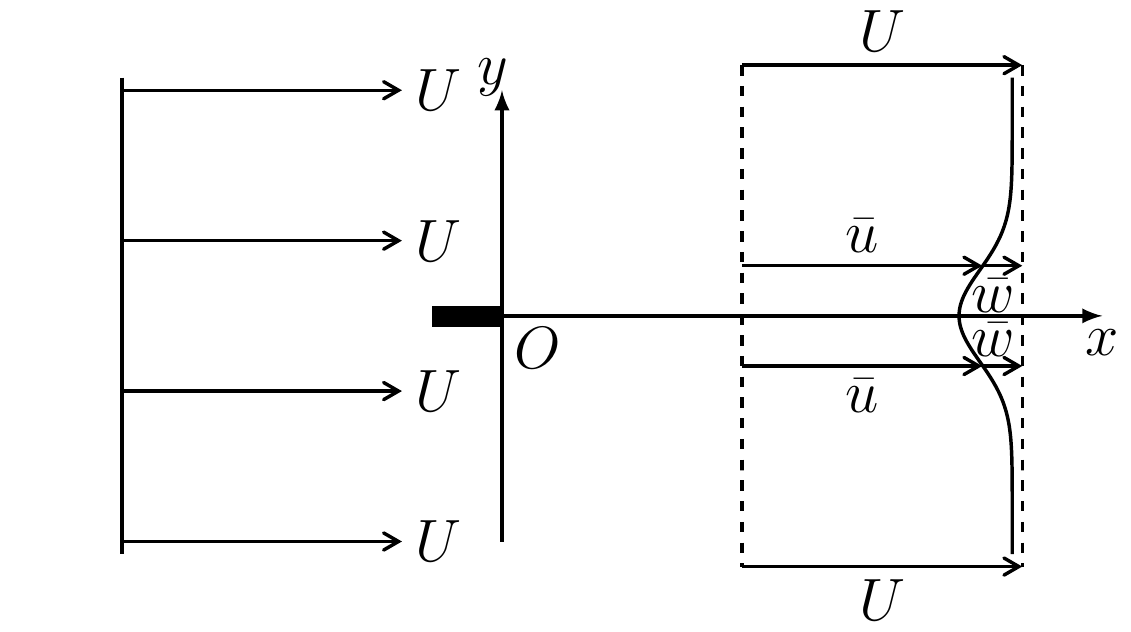}
    \caption{A two-dimensional turbulent classical wake behind a slender object $O$ aligned with the mainstream flow. The laminar mainstream flow has constant velocity $(U,0)$ and the mean velocity in the wake in the stream-wise direction is denoted by $\bar{u}$. In the far wake region, we define a mean velocity deficit $\bar{w}$.}\label{Wake}
\end{figure}

Consider a turbulent wake downstream of a slender stationary object which is referred to as a classical wake. We focus here on symmetric wakes which develop when the object is aligned with the laminar mainstream flow. We define the Cartesian coordinate system, $(x,y)$, so that the velocity of the mainstream flow is $(U,0)$, where $U$ is the constant mainstream speed. The origin of the coordinate system is placed at the trailing edge of the slender object. {Because} the object is slender, any length variation in the $y$-direction may be neglected, and we approximate its location as a finite line along a section of the negative $x$-axis. The fluid has constant density and dynamic \revtwo{(molecular)} viscosity, denoted by $\rho$ and $\mu$, respectively. The velocity components $(u,v)$ are decomposed into mean velocity components, $(\bar{u},\bar{v})$, and turbulent fluctuations, $({{u}^{\prime}},{{v}^{\prime}})$, so that $u = \bar{u} + {{u}^{\prime}}$ and $v = \bar{v} + {{v}^{\prime}}$. The pressure $p$ is also similarly decomposed into $\bar{p} + p'$. We assume that these mean quantities are independent of time. Flows of this kind are called \textit{steady turbulent flows}~\cite[p.~502]{schlichtingGersten}. {We consider the far downstream wake which behaves self-similarly.\footnote{\revtwo{The downstream distance at which self-similar behaviour is observed depends on the type of wake generator \cite{Wygnanski}.}}} In this region, a mean velocity deficit, $\bar{w}$, defined by $\bar{u} = U-\bar{w}$, is used to describe the flow \rev{and the inertia terms in the RANS equations can be linearised}. A turbulent wake is illustrated in Figure~\ref{Wake}.
\subsection{Governing equations}
To derive the governing equations for the mean velocity components in the turbulent far wake, the RANS equations are used as a starting point. These are obtained by substituting the flow variables $(\bar{u} + {{u}^{\prime}}, \bar{v} + {{v}^{\prime}},\bar{p}+p')$ into the Navier--Stokes equations, and then taking the time average. Although the time averages of the fluctuations are zero, averages of products of these fluctuations are nonzero \cite{Schlichting,P}. 

The $i$-$th$ component of the time-averaged momentum equation for steady turbulent flows is
\begin{equation}
\label{RANS}
    \rho \bar{u}_j \dfrac{\partial \bar{u}_i}{\partial x_j}= \dfrac{\partial}{\partial x_j} \left[-\bar{p} \delta_{i j} + \mu \left(\dfrac{\partial \bar{u}_i}{\partial x_j} +\dfrac{\partial \bar{u}_j}{\partial x_i}\right) - \rho \overline{u_i' u_j'}\right], \ \ i=1,2,
\end{equation}
where $u_1 = u$ and $u_2 = v$.
Three types of stresses arise: the isotropic stresses, $-\bar{p} \delta_{i j}$, the viscous stresses
\begin{equation}
  \bar{\tau}_{i j} = \mu \left(\dfrac{\partial \bar{u}_i}{\partial x_j} +\dfrac{\partial \bar{u}_j}{\partial x_i}\right),
\end{equation}
and the Reynolds or apparent stresses \cite{P}
\begin{equation}
\label{stressmeout}
    \bar{\tau}^T_{i j} = - \rho \overline{u_i' u_j'}.
\end{equation}
The Reynolds stresses are unknown resulting in more unknowns than equations, leading to the closure problem. To overcome the closure problem, these stresses 
must be specified. In the eddy viscosity model~\cite{Boussinesq1}, these Reynolds stresses are incorporated into the viscous stresses by defining a kinematic eddy viscosity $\nu_T = \mu_T/\rho$, where $\mu_T$ is the dynamic eddy viscosity, i.e.
\begin{equation}
\label{stressmain}
     \bar{\tau}^T_{i j} = \mu_T \left(\dfrac{\partial \bar{u}_i}{\partial x_j} +\dfrac{\partial \bar{u}_j}{\partial x_i}\right) - \dfrac{1}{3} \rho \overline{u_k' u_k'} \delta_{i j}.
\end{equation}

It is convenient to define an effective kinematic viscosity, $E$, which in algebraic closure models is taken as the sum of the kinematic viscosity and the turbulent or eddy kinematic viscosity, i.e. $E = \nu + \nu_T$. The form of $\nu_T$ (and hence $E$) depends on the flow geometry and also on the closure model used. Various closure models --- which in general may depend on $x$, $y$, $\bar{u}$, $\partial \bar{u}/\partial y$, and higher partial derivatives of $\bar{u}$ --- are discussed in Section~\ref{sec:closure}. A detailed derivation of the governing equations is given in \cite{ash}. For completeness, an outline is provided below.

We begin by defining some characteristic quantities. Let $L$ be the length in the $x$-direction beyond which the reduction in velocity is small enough to be neglected, and $E_{C}$ the characteristic effective {kinematic} viscosity.
The turbulent or modified Reynolds number is then 
\begin{equation}
Re_{T} = \frac{U L}{E_{C}},\label{prandrey}
\end{equation}
and is related to the Reynolds number, $Re = U L/\nu$, by 
\begin{equation}
    Re_{T} = Re \dfrac{\nu}{E_{C}}.
\end{equation}
A turbulent region downstream of the object can develop for large $Re$, but in order for a turbulent boundary layer to exist, terms of order $1/Re_T$ must be neglected~\cite{ash}. This is similar to the condition for a laminar boundary layer to exist, which requires terms of order $1/Re$ to be neglected~\cite{Schlichting}. Therefore, the turbulent Reynolds number must be large in order for the boundary layer approximation to be applied. In this work, we restrict our attention to large $Re_T$ and focus on fully developed turbulent boundary layers.

Introducing the dimensionless variables
\begin{equation}
x^* = \dfrac{x}{L},  \quad y^* = \dfrac{\sqrt{Re_T} }{L} y, \quad {{\bar{u}}}^* = \frac{{\bar{u}}}{U}, \quad {{\bar{v}}}^* = {\bar{v}}\frac{\sqrt{Re_T}}{U}, \quad  {\bar{p}}^* = \frac{p}{\rho U^2}, 
\label{dimvariables0}
\end{equation}
the boundary layer equations for the two-dimensional turbulent classical wake in terms of the dimensionless mean velocity components are~\cite{ash3} 
\begin{equation}\label{eq:Governing1}
    \pderiv{\bar{u}^*}{x^*}+\pderiv{\bar{v}^*}{y^*}=0,
\end{equation}
\begin{equation}\label{eq:Governing2}
    \bar{u}^*\pderiv{\bar{u}^*}{x^*}+\bar{v}^*\pderiv{\bar{u}^*}{y^*} = \pderiv{}{y^*}\left(E^*\pderiv{\bar{u}^*}{y^*}\right),
\end{equation}
where $E^*=E/E_C$ is the nondimensionalised effective {kinematic} viscosity.
The only surviving stress terms are the shear stresses
\begin{equation}
\label{stressbound}
  \bar{\tau}^*_{ x^* y^*} + \bar{\tau}^{*T}_{x^* y^*}  = \dfrac{1}{\sqrt{Re_T}}E^*\pderiv{\bar{u}^*}{y^*},
\end{equation}
where $\rho U^2$ is used to scale the shear stresses. Note that it is the $y^*$-derivative of this expression that appears in \Eref{eq:Governing2}, and that this term is of $\mathcal{O}(1)$.

As a consequence of choosing $U$ to nondimensionalise $\bar{u}$, the mainstream velocity is now unity in the $x^*$-direction. That is, far downstream from the body,
\begin{equation}
\bar{u}^*(x^*,y^*) = 1 - \bar{w}^*(x^*,y^*), \label{eq:deficitdef}
\end{equation}
where $\bar{w}^*(x^*,y^*)$ is the dimensionless mean velocity deficit in the $x^*$-direction. 
In the far wake region, $|\bar{w}^*| \ll \revone{1}$ and $|\bar{v}^*| \ll 1$, so products and powers of small terms can be neglected, and substituting { \Eref{eq:deficitdef}  into \Eref{eq:Governing1} and \Eref{eq:Governing2}} gives
\begin{equation}\label{eq:ConsofMass}
- \dfrac{\partial \bar{w}}{\partial x} + \dfrac{\partial \bar{v}}{\partial y} = 0, 
\end{equation}
\begin{equation}
\dfrac{\partial \bar{w}}{\partial x} = \dfrac{\partial}{\partial y} \left(E  \dfrac{\partial \bar{w} }{\partial y}  \right), \label{eq:finalequation}
\end{equation}
where the stars $^*$ have been suppressed for convenience.
Similarly, the shear stress terms (\ref{stressbound}) become
\begin{equation}
\label{stressbounda}
    \bar{\tau}_{ x y} + \bar{\tau}^{T}_{x y} = -\dfrac{1}{\sqrt{Re_T}}E\pderiv{\bar{w}}{y}.
\end{equation}

\subsection{Boundary conditions and the conserved quantity}

The momentum and conservation of mass equations, (\ref{eq:finalequation}) and (\ref{eq:ConsofMass}), must be solved subject to appropriate boundary conditions.
Consider first the mean velocity deficit $\bar{w}$ in the $x$-direction. When using the boundary layer theory approximation for the turbulent wake region, the \revone{effective} viscosity is neglected everywhere except in the \rev{shear} layer~\cite{Schlichting}. Boundary conditions are obtained by ensuring a smooth transition from the wake region to the inviscid mainstream flow at the boundary of the wake, which we denote by $\pm y_b(x)$. Although a finite wake boundary is nonphysical, we shall see later that $y_b(x)$ may be finite, or infinite, depending on the closure model used and in particular whether the kinematic viscosity is neglected or not. As a consequence of studying symmetric wakes, we restrict our attention to the upper half of the wake only, i.e. $y \geq 0$.

Mainstream matching provides two conditions. First, as $y$ tends to $ y_{b}(x)$ the mean velocity $\bar{u}$ tends to unity, and therefore the mean velocity deficit $\bar{w}$ will tend to zero, i.e. 
\begin{equation}
\bar{w} (x, y_{b}(x)) = 0.  \label{eq:bcdwinf}
\end{equation}
The second matching condition is given by Hutchinson \cite{AshVorticity}. At the boundary of the wake, $y= y_{b}(x)$, the {mean} vorticity, $\bar{\omega}$, must vanish to match that of the inviscid mainstream flow. In the boundary layer approximation we have
\begin{equation}
    \mathbf{\bar{\omega}} = \pderiv{\bar{u}}{y} = -\pderiv{\bar{w}}{y},
\end{equation}
and hence, 
\begin{equation}\label{eq:bcdwyinf}
 \dfrac{\partial \bar{w}}{\partial y} (x,y_{b}(x)) = 0.
\end{equation}
A further condition on $\bar{w}$ is imposed by the fact that the mean velocity deficit $\bar{w} (x,y)$ is a maximum with respect to $y$ at each point on the positive $x$-axis, \revone{which must hold for wakes symmetric about the $x$ axis}. Therefore,
\begin{equation}
\dfrac{\partial \bar{w}}{\partial y} (x,0) = 0, \quad x > 0.  \label{eq:bcdwzero}
\end{equation}

As the governing equations and boundary conditions are homogeneous (and $y_b(x)$ is unknown), an extra condition is required to complete the solution. This condition comes from a conserved quantity, which for the classical wake is the drag force \cite{Goldstein1933}. 
The conserved quantity imposes the constraint~\cite{ash3}
\begin{equation}\label{eq:ConservedQuant}
    \int^{y_{b}(x)}_{0} \bar{w}(x,y) dy = \dfrac{D}{2},
\end{equation}
where the \revtwo{dimensionless drag force per unit breadth} $D$ is independent of $x$.

Consider now $\bar{v}$, the mean velocity component in the $y$-direction. By the symmetry condition, $\bar{v}$ is zero along the positive $x$-axis, i.e. 
\begin{equation}
\bar{v} (x,0) = 0, \quad x > 0. \label{eq:bcdv}
\end{equation}
To derive an expression for $\bar{v}$, we substitute (\ref{eq:finalequation}) into (\ref{eq:ConsofMass}) and integrate with respect to $y$ to {obtain}
\begin{equation}
    \bar{v}(x,y) =E  \dfrac{\partial \bar{w} }{\partial y} + A(x),
\end{equation}
where $A(x)$ is an arbitrary function of $x$. If $E$ remains finite at $y=0$ then it follows from (\ref{eq:bcdwzero}) and (\ref{eq:bcdv}) that $A(x) \equiv 0$, and therefore,
\begin{equation}
\label{vcomp}
    \bar{v}(x,y) =E  \dfrac{\partial \bar{w} }{\partial y}.
\end{equation}
Furthermore, it follows from (\ref{eq:bcdwyinf}) that if $E$ remains finite as $y \rightarrow y_b(x)$ then $\bar{v}(x,y_b(x))=0$. Hence, if the effective viscosity is finite both on the centre line and the wake boundary, then \revone{this theoretical model predicts that there is no fluid entrainment. Large scale turbulent motions, which are not a feature of this model, are responsible for entrainment \cite{cantwell_1978}.}
The importance of $\bar{v}$ can be seen from \Cref{stressbounda} and \Cref{vcomp}, in that we can write
\begin{equation}
\label{shearnew}
    -\dfrac{1}{\sqrt{Re_T}} \bar{v}(x,y) = -\dfrac{1}{\sqrt{Re_T}} E  \dfrac{\partial \bar{w} }{\partial y} = \bar{\tau}_{x y}+\bar{\tau}^T_{x y},
\end{equation}
and find the shear stress can be determined by solving for $\bar{v}$.

\section{Derivation of the eddy viscosity for the EPML model}\label{sec:closure}

Application of the RANS equations to turbulent wake flows requires an appropriate closure model to complete the system~\Eref{eq:ConsofMass}--\Eref{eq:finalequation}~\cite{Boussinesq1}. 
In algebraic closure models, the effective kinematic viscosity is expressed as the sum of the kinematic viscosity $\nu$ and the kinematic eddy viscosity $\nu_{T}$ \cite{P}:
\begin{equation}
E = \frac{\mu + \mu_{T}}{\rho} =\nu + \nu_{T}. \label{eddyvis}
\end{equation}
Introducing nondimensional variables, 
\begin{equation} E^* = \frac{E}{E_{C}}, \quad \nu_T^* = \dfrac{\nu_T}{\nu_{TC}}, \label{dimvariablesa}
\end{equation}
where $E_{C}=\nu + \nu_{TC}$ and $\nu_{TC}$ is the characteristic kinematic eddy viscosity, the dimensionless effective viscosity, $E^*$, is given by
\begin{equation}
E^* = \nuchar + \nuTchar \nu^*_{T} \label{dimr}.
\end{equation}

Previous work investigates the class of models that can be described by a kinematic eddy viscosity of the form $\nu_T=\nu_T \left(x,y,\partial\bar{u}/\partial y\right)$  \cite{ash,ash2,ash3,ash4}.
In this paper, we extend the work conducted in \cite{ash3} by considering effective kinematic viscosities of the form 
\begin{equation}
E=E \left(x, \pderiv{\bar{u}}{y}, \pderiv[2]{\bar{u}}{y} \right).
\end{equation}
This form is convenient not only in that it extends the range of closure models that can be applied, but  it also incorporates the  CEV model, the PML model, the EPML model, and any variations thereof, as special cases, allowing for a unified derivation and a direct comparison.

Note that the boundary conditions \Eref{eq:bcdwinf}, \Eref{eq:bcdwyinf}, and \Eref{eq:bcdwzero} are all independent of the closure model used to define the effective viscosity. The only condition that must be imposed directly on the closure models  is that $\nu_T$ is finite at $y=0$ and at $y=y_b(x)$, which shows that no fluid entrainment occurs. Because a finite-valued effective viscosity on the entire domain is required to describe a turbulent flow --- large effective viscosities would decrease the turbulent Reynolds number $Re_T$, and the flow would no longer satisfy the condition for the existence of a turbulent boundary layer --- this seems reasonable.

\revone{For free shear flows, a viscous superlayer separates the turbulent flow region from the mainstream flow \cite{P}. However, this layer is thin compared to the boundary layer thickness $\delta$ and so the wake boundary is essentially a well-defined interface. }
Since the wake boundary is where the turbulent wake merges with the laminar mainstream flow, the eddy viscosity $\nu_T$ should vanish there. Consider then the momentum equation (\ref{eq:finalequation}) with an effective viscosity of the form~(\ref{dimr}). If $\nu_T \rightarrow 0$ as $y \rightarrow \pm y_b(x)$ then for $y$ sufficiently close to $\pm y_b(x)$, $\nu \gg \nu_T$. 
Hence, as $y \rightarrow \pm y_b(x)$, \Cref{eq:finalequation} becomes
\begin{equation}
\label{exponential}
\dfrac{\partial \bar{w}}{\partial x} = {\dfrac{\nu}{E_C}} \dfrac{\partial^2 \bar{w}}{\partial y^2},
\end{equation}
and the exponential solution for a laminar wake will apply~\cite{Goldstein1933}. Therefore, $\bar{w}$ can not reach zero for any finite value of $y_b$ and we must have that $y_b(x) = \infty$. A finite wake can therefore only exist if $\nu = 0$ everywhere in the approximation. We shall see later  that the condition $\nu_T \rightarrow 0$ as $y \rightarrow \pm y_b(x)$ is satisfied for the PML and EPML models, but not for the CEV model.

\subsection{Prandtl's extended mixing length model}
In an attempt to improve upon the accuracy of the results obtained when the constant eddy viscosity model is applied, Prandtl introduced a mixing length model (PML).\footnote{A detailed description of the origins of the ideas pertaining to the PML model can be found in \cite{Brad}.} \rev{Prandtl then extended this model to address some of its shortcomings \cite{Prandtl}. However, to the best of the authors' knowledge, the extended model (EPML)  was not accompanied by a detailed explanation or derivation.
We now present a derivation of the EPML model, which relies on an adapted derivation of the PML model for parallel mean flow as provided by Schlichting and Schlichting \& Gersten \cite{Schlichting,schlichtingGersten}.}

\begin{figure}
    \centering
    \includegraphics[scale=0.5]{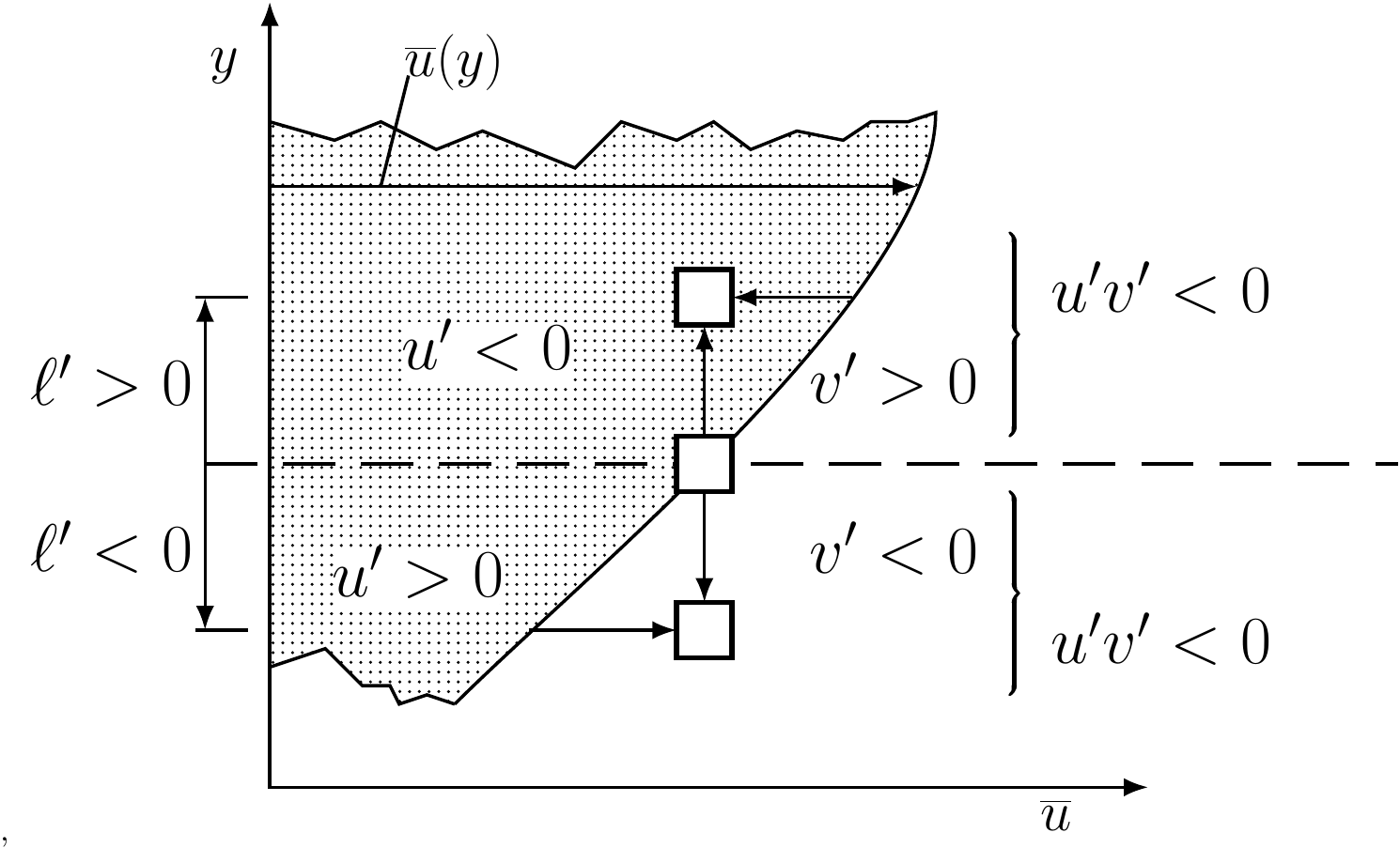}
    \caption{Explanation of the mixing length concept (diagram adapted from Schlichting \cite[p 539]{schlichtingGersten}). Fluid lumps at a position $y$ are displaced by random turbulent fluctuations to position $y + {{\ell}^{\prime}}$. These lumps retain their momentum, and hence velocity. The mean velocity of the fluid is denoted by $(\bar{u}(y),0)$ and the turbulent velocity fluctuations by $({{u}^{\prime}},{{v}^{\prime}})$. The difference in mean velocity of the fluid lumps and that of the surrounding fluid is used to estimate the strength of the turbulent fluctuations. Note that, as a result of conservation of mass, ${{u}^{\prime}} {{v}^{\prime}}<0$.}  
    \label{fig:PrandtlDiagram}
\end{figure}

Consider a turbulent flow with parallel mean flow as shown in \Cref{fig:PrandtlDiagram}. In turbulent flows, fluid particles coalesce forming lumps of fluid, which then travel as a whole in both the $x$ and $y$ directions. These lumps of fluid remain intact and retain their momentum, travelling some distance before mixing in with the surrounding fluid once again.  In parallel laminar flows, the $y$-component of the velocity is zero and so fluid elements cannot travel in the transverse direction. However, in parallel turbulent flows, turbulent fluctuations displace fluid lumps in the transverse direction. These displacements are random.  The standard deviation, (with a multiplicative constant that can be absorbed), of these displacements is known as the {\em mixing length}, from which this model derives its name.

In the turbulent flow shown in \Cref{fig:PrandtlDiagram},
the parallel mean velocity is denoted by $(\bar{u}(y),0)$ and the random velocity fluctuations by $(u^\prime,v^\prime)$. The fluid velocity is then
\begin{equation}
\label{head}
    (u(x,y,t),v(x,y,t)) = (\bar{u}(y) + u^\prime(x,y,t),v^\prime(x,y,t)).
\end{equation}
Consider a mean flow where $\mathrm{d}\bar{u}/\mathrm{d}y>0$. 
Fluid lumps are displaced by the turbulent fluctuations from position $y$ to position $y+\ell^\prime$, whilst retaining their original momentum in the $x$-direction.\footnote{\rev{In \cite{Schlichting}, turbulent fluctuations are caused by fluid lumps arriving at a layer from layers above and below, and in \cite{schlichtingGersten}, fluctuations cause fluid lumps to leave a layer and move into a neighbouring layer.}} In general, the random variable $\ell^\prime$ is a function of $x$ and $y$ and time $t$, and can take on both positive and negative values.  For the purposes of clear illustration we will consider positive $\ell^\prime$ values, and then later relax this condition.
The size of $\ell^\prime$ gives an indication as to the strength of the fluctuations that resulted in a fluid lump moving from $y$ to $y+\ell^\prime$. 
Because the fluid lumps retain their momentum and hence velocity in the $x$-direction while being displaced, they arrive at the new layer $y+\ell^\prime$ with a lower velocity of $\bar{u}(y)$ than that of the surrounding fluid, $\bar{u}(y+\ell^\prime)$.

To estimate the strength of these fluctuations, we use the difference between the velocity of the surrounding fluid and the newly arrived fluid lump:
\begin{equation}\label{eqn:triu}
    \triangle u = \bar{u}(y+\ell^\prime)-\bar{u}(y).
\end{equation}
\rev{Assuming that ${{\ell}^{\prime}}$ is small, expanding $\bar{u}(y+\ell^\prime)$ as a Taylor series gives
\begin{equation}
\label{eq:uprimeprandtl}
    \triangle u = {{\ell}^{\prime}} \dfrac{d\bar{u}}{dy} + \dfrac{{{\ell}^{\prime}}^2}{2}\dfrac{d^2 \bar{u}}{d y^2} +\dfrac{{{\ell}^{\prime}}^3}{6}\dfrac{d^3 \bar{u}}{d y^3}+\dfrac{{{\ell}^{\prime}}^4}{24}\dfrac{d^4 \bar{u}}{d y^4}+ \mathcal{O}({{{\ell}^{\prime}}^5}).
\end{equation}
In the PML model, only the first term in \Cref{eq:uprimeprandtl} is retained. However, for points where $d \bar{u}/dy = 0$,  the first term in~(\ref{eq:uprimeprandtl}) vanishes, and $\triangle u$ is zero if higher order terms are neglected. In the EPML model, points at which $d \bar{u}/dy = 0$ are taken into account, and additional terms must be included. In particular, terms up to and including ${\cal{O}}({{{\ell}^{\prime}}^4})$ are retained in the EPML model.}

\rev{Before we can express the fluctuations ${{u}^{\prime}}$ and ${{v}^{\prime}}$ in terms of $\triangle u$, we must first determine their signs and order of magnitude.} \rev{In order for a fluid lump to be displaced in the positive $y$-direction, the velocity fluctuation, ${{v}^{\prime}}$, must be positive. Furthermore, because a fluid lump  initially at $y$ which is displaced to $y+\ell^{\prime}$ has a lower velocity than that of its surroundings, ${{u}^{\prime}}$ must be negative (see Figure~\ref{fig:PrandtlDiagram}). Therefore, ${{u}^{\prime}}$ and ${{v}^{\prime}}$ must have opposite signs.}
\rev{For small-scale turbulent motions (the type which are considered here), it is reasonable to assume that characteristic length scales in the $x$ and $y$ directions are comparable.\footnote{\revone{This assumption may not hold for large-scale motions.}} By comparing the magnitude of the terms in the conservation of mass equation,
\begin{equation}
    \frac{\partial {{u}^{\prime}}}{\partial x} + \frac{\partial {{v}^{\prime}}}{\partial y} = 0,
\end{equation}
we find that the fluctuations in the $x$ and $y$ directions must also be of the same magnitude, i.e. $|{{v}^{\prime}}| \sim  |{{u}^{\prime}}|$~\cite{Schlichting}.}
\rev{We may therefore write
\begin{equation}
  {{v}^{\prime}} = -c {{u}^{\prime}}>0,
\end{equation}
where $c$ is a positive constant of order $1$.}

\rev{Recalling that, for the moment, we are considering $\ell^{\prime} > 0$ and $d \bar{u}/d y>0$,  the leading order term in~(\ref{eq:uprimeprandtl}) is positive and hence, for sufficiently small $\ell^{\prime}$ we have $\triangle u>0$. Now, since $\triangle u$ is an estimate of the difference in mean velocity between a fluid lump and its surroundings (see~(\ref{eqn:triu})),  we assume}
\rev{
\begin{equation}
  {{u}^{\prime}} = -\triangle u,
\end{equation}
where the minus comes from comparing the signs of $u^\prime$ and $\triangle u$ as described above.}

\rev{Using the Reynolds decomposition, we may write the $x$-component of the velocity, $u$, at position $(x,y)$ and time $t$ as
\begin{equation}
 u(x,y,t) = \bar{u}(y) + {{u}^{\prime}} = \bar{u}(y)  - \ell^\prime \deriv{\bar{u}}{y} -  \dfrac{{{\ell}^{\prime}}^2}{2}\dfrac{d^2 \bar{u}}{d y^2} - \dfrac{{{\ell}^{\prime}}^3}{6}\dfrac{d^3 \bar{u}}{d y^3}-\dfrac{{{\ell}^{\prime}}^4}{24}\dfrac{d^4 \bar{u}}{d y^4},   
\end{equation}
and the $y$-component as
\begin{equation}    
v(x,y,t) = v^\prime = -c {{u}^{\prime}} = + c \left({{\ell}^{\prime}} \deriv{\bar{u}}{y}  + \dfrac{{{\ell}^{\prime}}^2}{2}\dfrac{d^2 \bar{u}}{d y^2} +\dfrac{{{\ell}^{\prime}}^3}{6}\dfrac{d^3 \bar{u}}{d y^3}+\dfrac{{{{\ell}^{\prime}}^4}}{24}\dfrac{d^4 \bar{u}}{d y^4}\right).
\end{equation}
Allowing ${{\ell}^{\prime}}$ to now take on both positive and negative values and averaging over time (see~(\ref{stressmeout})), the Reynolds shear stress, $\bar{\tau}_{x y}^T$, is given by
\begin{equation}
 \bar{\tau}^T_{xy} = -\rho \overline{{{u}^{\prime}} {{v}^{\prime}}} = \rho c \left[\overline{{{{\ell}^{\prime}}^{2}}} \left(\dfrac{d \bar{u}}{d y} \right)^2 + \overline{{{{\ell}^{\prime}}^3}} \dfrac{d \bar{u}}{d y}\dfrac{d^2 \bar{u}}{d y^2} + \dfrac{\overline{{{{\ell}^{\prime}}^4}}}{4} \left(\dfrac{d^2 \bar{u}}{d y^2} \right)^2 + \dfrac{\overline{{{{\ell}^{\prime}}^4}}}{3} \dfrac{d \bar{u}}{d y}\dfrac{d^3 \bar{u}}{d y^3} \right] + \mathcal{O}(\overline{{{\ell}^{\prime}}^5}).
\end{equation}}

\rev{The EPML model specifically aims to improve on the approximation used by the PML model by taking into account points at which $d \bar{u}/dy$ vanishes. Therefore, we retain only the lowest order term and those higher order terms that do not depend on $d \bar{u}/dy$, giving
\begin{equation}
 \bar{\tau}^T_{xy} = -\rho \overline{{{u}^{\prime}} {{v}^{\prime}}} = \rho c \left[\overline{{{{\ell}^{\prime}}^2}} \left(\dfrac{d \bar{u}}{d y} \right)^2 + \dfrac{\overline{{{\ell}^{\prime}}^4}}{4} \left(\dfrac{d^2 \bar{u}}{d y^2} \right)^2  \right] + \mathcal{O}(\overline{{{\ell}^{\prime}}^5}).
\end{equation}
Setting $\ell_1^2= c \overline{{{{\ell}^{\prime}}^2}}$, $\ell_2^2 = \overline{{{\ell}^{\prime}}^4}/ \overline{ {{\ell}^{\prime}}^2}$, and neglecting terms of $\mathcal{O}(\overline{{{\ell}^{\prime}}^5})$  leads to
\begin{equation}
\label{Newtau}
 \bar{\tau}^T_{xy} = \rho  \ell_1^2 \left[ \left(\dfrac{d \bar{u}}{d y} \right)^2 + \dfrac{\ell_2^2}{4} \left(\dfrac{d^2 \bar{u}}{d y^2} \right)^2  \right],
\end{equation}
where the first mixing length, $\ell_1$, is simply a constant multiple of the standard deviation, and the second mixing length, $\ell_2$, is the kurtosis.} 

\rev{
We will consider flows where $d \bar{u}/d y>0$ almost everywhere, except at a finite number of points where $d \bar{u}/d y=0$. Therefore, over the majority of the flow domain,
\begin{equation}
\label{maj}
   \left(\dfrac{d \bar{u}}{d y} \right)^2 \gg \ell_2^2 \left(\dfrac{d^2 \bar{u}}{d y^2} \right)^2,
\end{equation}
which is a reasonable assumption when $\ell_2$ is small compared to the boundary layer thickness, $\ell_2/\delta \ll 1$. 
A reformulated version of the EPML model using this concept has been applied to turbulent pipe flows \cite{Note} and flows in circular tubes~\cite{Luo}. However, to obtain Prandtl's version of the extended mixing length model, which can be written in the form~\cite{Prandtl},
\begin{equation}
\label{PrandV}
    \bar{\tau}^T_{xy}=  \rho \ell_1^2 \sqrt{\left(\deriv{\bar{u}}{y}\right)^2 + \dfrac{\ell_2^2}{2}\left(\deriv[2]{\bar{u}}{y}\right)^2}\dfrac{d \bar{u}}{d y},
\end{equation}
we simply note that when~(\ref{maj}) holds, substituting the binomial approximation 
\begin{equation}
   \sqrt{\left(\deriv{\bar{u}}{y}\right)^2 + \dfrac{\ell_2^2}{2}\left(\deriv[2]{\bar{u}}{y}\right)^2} \approx \dfrac{d \bar{u}}{d y}  + \dfrac{\ell_2^2}{4 \deriv{\bar{u}}{y} }\left(\deriv[2]{\bar{u}}{y}\right)^2,
\end{equation}
into~(\ref{PrandV}), results in (\ref{Newtau}).}

Looking now at \Cref{stressmain}, the normal stresses,   $\bar{\tau}^T_{xx}$ and $\bar{\tau}^T_{yy}$, vanish (a consequence of the mean parallel flow), and so the only remaining term is the shear stress
\begin{equation}\label{eq:prandtlmixinglengthtau}
    \bar{\tau}^T_{xy} = \mu_T \deriv{\bar{u}}{y}.
\end{equation}
Comparing this expression to (\ref{PrandV}) leads to
\begin{equation}
\label{mudim}
    \mu_T = \rho \ell_1^2 \sqrt{\left(\deriv{\bar{u}}{y}\right)^2 + \ell_2^2\left(\deriv[2]{\bar{u}}{y}\right)^2},
\end{equation}
where the constant factor of $1/2$ has been absorbed into $\ell_2$ for convenience. This result also holds for the case where $d \bar{u}/d y<0$.

In boundary layer flows,
\begin{equation}
    \left| \dfrac{\partial \bar{v}}{\partial x} \right| \ll \left| \dfrac{\partial \bar{u}}{\partial y} \right|,
\end{equation}
and the Reynolds stress, $\bar{\tau}^T_{x y}$, depends only on $\partial \bar{u}/\partial y$  as shown in \Cref{stressbound}.
Therefore, the result in \Cref{mudim} can be extended to boundary layers. For wakes, we replace the ordinary derivative $\mathrm{d} \bar{u}/\mathrm{d}y$ with $-\partial \bar{w}/\partial y$:
\begin{equation}
\label{mudim2}
\nu_T =    \dfrac{\mu_T}{\rho} =  \ell_1^2 \sqrt{\left(\pderiv{\bar{w}}{y}\right)^2 + \ell_2^2\left(\pderiv[2]{\bar{w}}{y}\right)^2}.
\end{equation}
In Prandtl's mixing length model for free shear flows,  the mixing length is a function of $x$ alone.
Define dimensionless lengths 
\begin{equation}
    \ell_1^*(x^*) = \dfrac{\ell_1(x)}{\delta}, \ \ \ \ell_2^*(x^*) = \dfrac{\ell_2(x)}{\delta},
\end{equation}
where $\delta =  L/\sqrt{Re_T}$ is the boundary layer thickness.
In dimensionless form,  \Cref{mudim2} is
\begin{equation}
    \nu_T^*=  {\ell_{1}^*}^2(x^*)\sqrt{\left(\pderiv{\bar{w}^*}{y^*}\right)^2+ {\ell_{2}^*}^2(x^*)\left(\pderiv[2]{\bar{w}^*}{y^*}\right)^2},
\end{equation}
and the effective viscosity, \Cref{dimr}, is 
\begin{equation}\label{eq:New5.144}
 E^*=\nuchar+\nuTchar  {\ell_{1}^*}^2(x^*)\sqrt{\left(\pderiv{\bar{w}^*}{y^*}\right)^2+ {\ell_{2}^*}^2(x^*)\left(\pderiv[2]{\bar{w}^*}{y^*}\right)^2},   
\end{equation}
where $\nu_{TC} = U \delta$.

To simplify the notation, let
\begin{equation}\label{eq:New5.145}
 \beta = \nuchar, \quad \alpha = \nuTchar. 
\end{equation}
In terms of (\ref{eq:New5.145}), \Cref{eq:New5.144} is 
\begin{equation}\label{eq:New5.146}
    E\left(x,\pderiv{w}{y},\pderiv[2]{w}{y}\right) = \beta + \alpha \ell_1^2(x) \left[\left(\pderiv{w}{y}\right)^2 + \ell_2^2(x)\left(\pderiv[2]{w}{y}\right)^2\right]^{1/2},
\end{equation}
where the bars and stars have been suppressed for further convenience. \Cref{eq:finalequation}  with the effective viscosity defined by  \Cref{eq:New5.146} is then
\begin{equation}\label{eq:IEPM1}
    \pderiv{w}{x}=\pderiv{}{y}\left[\beta \pderiv{w}{y} + \alpha \ell_1^2(x) \left[\left(\pderiv{w}{y}\right)^2+\ell_2^2(x)\left(\pderiv[2]{w}{y}
    \right)^2\right]^{1/2}\pderiv{w}{y}\right].
\end{equation}

\rev{The PML model is a special case of the extended model that is obtained when $\beta$ is set to zero and the term containing the second mixing length $\ell_2$ is neglected. Prandtl excluded the kinematic viscosity  on the basis that it is much smaller than the turbulent viscosity.  However, from (\ref{eq:bcdwyinf}) and (\ref{eq:bcdwzero}), the eddy viscosity vanishes at the wake boundary and on the wake axis, and this is no longer true. }
It was shown in \cite{ash} that including the kinematic viscosity leads to an infinite wake boundary, and the form of the mixing length can be obtained without imposing additional restrictions. \revtwo{Whilst idealised boundary conditions on unbounded domains may seem a long way from realistic experiments, we shall see in Section \ref{sec5} that in practice the computed velocity profiles decay rapidly -- exponentially in some cases -- indicating that the assumption of an unbounded domain has little effect on the behaviour model.}

In the next section, scaling solutions will be investigated for the case where the kinematic viscosity is neglected ($\beta=0$), and for when it is included ($\beta>0$) for the EPML model. \rev{In Section~\ref{sec5} we show that the inclusion of the second mixing length in the EPML model can significantly improve upon the PML model when comparing the resulting mean velocity profiles to experimental data.}

\section{Scaling solutions}\label{sec4}
We now investigate when the PDE (\ref{eq:IEPM1}) is invariant under the scaling transformation
\begin{equation}\label{eq:IEPM2v2}
    x = \lambda^{a} \bar{x}, \quad y = \lambda^{b} \bar{y}, \quad w = \lambda^{c} \bar{w}, \quad \ell_1 = \lambda^{d} \bar{\ell}_1, \quad \ell_2 = \lambda^{e} \bar{\ell}_2,
\end{equation}
and thereby reduce \Cref{eq:IEPM1} to an ODE. In the scaling transformation (\ref{eq:IEPM2v2}), only the ratios of $a, b, c, d,$ and $e$ need to be determined. If $a \neq 0$, we therefore need to determine only the four ratios $b/a, c/a, d/a,$ and $e/a$ and only four conditions need to be found. Without loss of generality, we set $a=1$. The boundary conditions are invariant under the transformation (\ref{eq:IEPM2v2}) because they are homogeneous. Therefore, the conditions for invariance are obtained from the equation itself, (\ref{eq:IEPM1}), and the conserved quantity.

We first examine the case in which the kinematic viscosity $\nu$ is neglected as considered by Prandtl. We will see that only three conditions are obtained and to determine the scaling transformation completely, one additional condition needs to be imposed. We will impose Prandtl's hypothesis~\cite{PrandtlHypothesis}. We then extend this work and find the scaling transformation when $\nu$ is included in \Cref{eq:IEPM1}. We find that four conditions are obtained and the scaling transformation is completely determined. An additional condition is not required. Analytical and numerical solutions of the ODE obtained in the reduction, subject to the boundary conditions and the conserved quantity, are derived.

\subsection{Extended Prandtl model without kinematic viscosity}

In Prandtl's original version of the extended mixing length model, the kinematic viscosity is neglected. The eddy viscosity for this model is a special case of \Cref{eq:New5.146} with $\beta=0$.
\Cref{eq:IEPM1} becomes 
\begin{equation}\label{eq:IEPM1v2}
    \pderiv{w}{x}=\alpha \ell_1^2(x)\pderiv{}{y}\left[ \left[\left(\pderiv{w}{y}\right)^2+\ell_2^2(x)\left(\pderiv[2]{w}{y}
    \right)^2\right]^{1/2}\pderiv{w}{y}\right].
\end{equation}
In terms of the scalings defined in \Cref{eq:IEPM2v2} with $a=1$, \Cref{eq:IEPM1v2} becomes
\begin{equation}\label{eq:IEPM3v2}
    \pderiv{\bar{w}}{\bar{x}} = \lambda^{1-3b+c+2d}\alpha \bar{\ell}_1^2\pderiv{}{\bar{y}}\left[  \left[ \left(\pderiv{\bar{w}}{\bar{y}}\right)^2+\lambda^{-2b+2e} \bar{\ell}_2^2 \left(\pderiv[2]{\bar{w}}{\bar{y}}\right)^2\right]^{1/2}\pderiv{\bar{w}}{\bar{y}}\right],
\end{equation}
and the conserved quantity (\ref{eq:ConservedQuant}) is given by 
\begin{equation}\label{eq:IEPM4v2}
    \dfrac{D}{2}=\lambda^{c+b}\int_0^{\bar{y}_b} \bar{w} (\bar{x}, \bar{y}) d\bar{y},
\end{equation}
where $ \bar{y}_b (\bar{x}) = y_b(x)/\lambda^b$.
Now, Equations~(\ref{eq:IEPM3v2}) and~(\ref{eq:IEPM4v2}) are invariant under the transformation (\ref{eq:IEPM2v2}) provided
\begin{equation}
    1-3 b + c + 2 d = 0, \ \ b - e = 0, \ \ b + c = 0,
\end{equation}
that is, provided
\begin{equation}\label{eq:PrandtlExtendedPartialScaling}
     c=-b, \quad d=\dfrac{1}{2} \left(4 b -1 \right), \quad e=b.
\end{equation}
Only three conditions have been obtained relating the four unknowns. One further condition is required. A condition that we will consider is Prandtl's hypothesis which states that the mixing length is proportional to the width of the boundary \revone{layer} \cite{PrandtlHypothesis}. 
In addition to Prandtl's hypothesis, we consider two other cases.

\textit{Case $1$}: $\ell_1(x) \propto y_{b}(x)$. Imposing Prandtl's hypothesis on the first mixing length $\ell_1(x)$ gives
\begin{equation}\label{eq:ell1propboundary}
    \ell_1(x) = k_1 y_{b}(x),
\end{equation}
where $k_1$ is a constant of proportionality. In terms of the scaled variables,
\begin{equation}\label{eq:ell1propboundaryv2}
    \bar{\ell}_1 = \lambda^{b-d}  k_1  \bar{y}_{b},
\end{equation}
and we see that $d=b$ for invariance. Therefore, this condition, combined with the ones in \Cref{eq:PrandtlExtendedPartialScaling} results in 
\begin{equation}\label{eq:PrandtlExtendedInvarianceConditions}
    b = \dfrac{1}{2}, \quad c=-\dfrac{1}{2}, \quad d=\dfrac{1}{2}, \quad e=\dfrac{1}{2},
\end{equation}
and the scaling transformation (\ref{eq:IEPM2v2}), which is now completely determined, takes the form
\begin{equation}
\label{MainScale}
   x = \lambda \bar{x}, \quad y = \lambda^{1/2} \bar{y}, \quad w = \lambda^{-1/2} \bar{w}, \quad \ell_1 = \lambda^{1/2} \bar{\ell}_1, \quad \ell_2 = \lambda^{1/2} \bar{\ell}_2. 
\end{equation}

\textit{Case $2$}: $\ell_2(x) \propto y_{b}(x)$. Now suppose that we apply Prandtl's hypothesis on the second mixing length, $\ell_2(x)$. Then
\begin{equation}\label{eq:ell2propboundary}
    \ell_2(x) = k_2 y_{b}(x),
\end{equation}
where $k_2$ is a constant of proportionality. In terms of the scaled variables we have
\begin{equation}\label{eq:ell2propboundaryv2}
    \bar{\ell}_2 = \lambda^{b-e} k_2  \bar{y}_{b}.
\end{equation}
However, from \Cref{eq:ell2propboundaryv2} we find $b=e$ for invariance which is a repeat condition and  we are still short of one condition. We see that Prandtl's hypothesis must be applied to the first mixing length.

\textit{Case $3$}: $\ell_1(x) \propto \ell_2(x)$. In this case, we have
\begin{equation}\label{eq:ell1propell2}
    \ell_1(x) = k_3 \ell_2(x),
\end{equation}
where $k_3$ is a constant of proportionality. 
In terms of the scaled variables, \Cref{eq:ell1propell2} becomes
\begin{equation}\label{eq:ell1propell2final}
    \bar{\ell}_1 = \lambda^{e-d} k_3 \bar{\ell}_2.
\end{equation}
From \Cref{eq:ell1propell2final} we see that $d=e$ for invariance and so using (\ref{eq:PrandtlExtendedPartialScaling}) we have $d=b$ which results in the same scaling as in \Cref{MainScale}.

Prandtl's hypothesis is satisfied when the mixing lengths are assumed to be proportional. Thus, Prandtl's hypothesis is verified for a special case of a two-dimensional turbulent classical wake described by Prandtl's extended model for the eddy viscosity in which the two mixing lengths are proportional.

\subsection{Extended Prandtl model with kinematic viscosity}

In this section we consider an improved version of the EPML model where the kinematic viscosity is included.  Here, the two mixing lengths are taken to be distinct and we will show that for an invariant solution to exist the two mixing lengths must be proportional. 

Consider again a scaling transformation given by \Cref{eq:IEPM2v2}. \Cref{eq:IEPM1} transforms to 
\begin{equation}\label{eq:IEPM3}
    \pderiv{\bar{w}}{\bar{x}} = \pderiv{}{\bar{y}}\left[\lambda^{1-2b}\beta \pderiv{\bar{w}}{\bar{y}}+ \lambda^{1-3b+c+2d} \alpha \bar{\ell}_1^2\left[ \left(\pderiv{\bar{w}}{\bar{y}}\right)^2+\lambda^{-2b+2e} \bar{\ell}_2^2 \left(\pderiv[2]{\bar{w}}{\bar{y}}\right)^2\right]^{1/2}\pderiv{\bar{w}}{\bar{y}}\right].
\end{equation}
Thus \Cref{eq:IEPM3} and the conserved quantity (\ref{eq:IEPM4v2}) are invariant provided 
\begin{equation}
    1 - 2 b = 0, \ \ 1 - 3 b + c + 2 d = 0, \ \ b - e = 0, \ \ b + c = 0.
\end{equation}
The additional condition, $1 - 2 b = 0$ is obtained. Hence, we again obtain the result in \Cref{eq:PrandtlExtendedInvarianceConditions}. Including the kinematic viscosity leads to the same result, (\ref{eq:PrandtlExtendedInvarianceConditions}), without the requirement that Prandtl's hypothesis be imposed as was done in Case 1 or the assumption that $\ell_1 (x) \propto \ell_2 (x)$ as was done in Case 3.

The additional condition that needed to be imposed for $\nu=0$ to  obtain the complete scaling solution can be formulated alternatively as follows: The solution for $\nu \neq 0$ must match with the solution for $\nu = 0$ in the limit as $\nu \rightarrow 0$. This again gives $b=1/2$. We can use this to replace Prandtl's hypothesis as these conditions are equivalent. We now see that because the original extended model with $\beta=0$ is a special case of the extended model with $\beta \neq 0$, we need only to consider the scaling solution in \Cref{MainScale} and reduce the PDE to an ODE. 
By using the method described in the text \cite{Dres}, it can be shown that the invariant solution under the scaling transformation \Eref{eq:IEPM2v2} with $a=1$ is of the form
\begin{equation}
    w(x,y) =x^c F(\xi), \ \ \ \ \xi = \dfrac{y}{x^b},
\end{equation}
\begin{equation}
    \ell_1(x) = K_1 x^d, \ \ \ell_2(x) = K_2 x^e,
\end{equation}
where $F(\xi)$ is an arbitrary function to be determined, and $K_1$ and $K_2$ are constants. Hence, from \Eref{eq:PrandtlExtendedInvarianceConditions}, the invariant solution is
\begin{equation}
\label{eq:SimVariablesa}
    w(x,y) = \dfrac{F(\xi)}{\sqrt{x}}, \ \ \ \ \xi = \dfrac{y}{\sqrt{x}},
\end{equation}
\begin{equation}\label{eq:IEPMMFinal}
    \ell_1(x) = K_1 \sqrt{x}, \ \ \ell_2(x) = K_2 \sqrt{x}.
\end{equation}
 This applies for Case 1 and Case 3 of the extended model with $\beta=0$ and for the extended model with $\beta \neq 0$. It is readily seen that $\ell_1 \propto \ell_2$ and therefore by including the kinematic viscosity, the two mixing lengths are found to be proportional without any additional hypothesis. \revone{From \Cref{eq:SimVariablesa}, we see that the invariant solution only applies for finite $x$. As $x \rightarrow \infty$, the velocity deficit tends to zero and the flow reverts to the undisturbed mainstream flow.}

Expressing \Cref{eq:IEPM1} in terms of the similarity variables (\ref{eq:SimVariablesa})--\Eref{eq:IEPMMFinal}  gives
\begin{equation}\label{eq:IEPM11}
  \deriv{}{\xi}\left[\beta \deriv{F}{\xi} + \alpha K_1^2 \deriv{F}{\xi}\left[\left(\deriv{F}{\xi}\right)^2 + K_2^2 \left(\deriv[2]{F}{\xi}\right)^2\right]^{1/2}  \right]+\dfrac{1}{2} \deriv{}{\xi}\left[\xi F\right]=0.  
\end{equation}
Note that the $y$-component of the velocity, \Cref{vcomp}, in terms of the similarity variables (\ref{eq:SimVariablesa}) - \Eref{eq:IEPMMFinal} is 
\begin{equation}
\label{ycompvel}
    v(x,y) = \dfrac{1}{x}\left( \beta \deriv{F}{\xi} + \alpha K_1^2 \deriv{F}{\xi}\left[\left(\deriv{F}{\xi}\right)^2 + K_2^2 \left(\deriv[2]{F}{\xi}\right)^2\right]^{1/2}\right),
\end{equation}
which can be written in the form
\begin{equation}
\label{vprof}
    v(x,y) = \dfrac{G(\xi)}{x},
\end{equation}
where
\begin{equation}
\label{G}
    G(\xi) =\beta \deriv{F}{\xi}+  \alpha K_1^2 \deriv{F}{\xi}\left[\left(\deriv{F}{\xi}\right)^2 + K_2^2 \left(\deriv[2]{F}{\xi}\right)^2\right]^{1/2}.
\end{equation}

The conserved quantity in \Cref{eq:ConservedQuant} transforms to 
\begin{equation}\label{eq:ConservedQuantityITOF}
    \int_0^{\dfrac{y_b(x)}{\sqrt{x}}} F d\xi = \dfrac{D}{2}.
\end{equation}
Now, $D$ is a constant, independent of $x$, provided
\begin{equation}
    \dfrac{y_b(x)}{\sqrt{x}}=\text{constant}=\xi_b.
\end{equation}
Hence, the boundary of the wake is given by
\begin{equation}
    y_b(x) = \xi_b \sqrt{x}, \quad \xi_b = \text{constant}.
\end{equation}
If the boundary of the wake extends to infinity \revone{at some finite distance downstream}, then $\xi_b = \infty$.
For $x > 0$, the boundary conditions in (\ref{eq:bcdwzero}) (\ref{eq:bcdwinf}), and (\ref{eq:bcdwyinf}) are now given in terms of the similarity variables by
\begin{equation}\label{eq:bcdFxib}
 F(\xi_{b}) = 0,  \ \  \deriv{F}{\xi}(\xi_{b}) =0, \ \  \deriv{F}{\xi}(0) =0,
\end{equation}
respectively.

\Cref{eq:IEPM11} is an example of the double reduction theorem of Sj{\"o}berg \cite{S} that if a PDE is reduced to an ODE by a symmetry associated with a conserved vector of the PDE, then the ODE can be integrated at least once. Integrating \Cref{eq:IEPM11} once yields
\begin{equation}\label{eq:IEPM12}
    \xi F + 2 \beta \deriv{F}{\xi} + 2 \alpha K_1^2 \deriv{F}{\xi}\left[\left(\deriv{F}{\xi}\right)^2 + K_2^2 \left(\deriv[2]{F}{\xi}\right)^2\right]^{1/2}   = a_1,
\end{equation}
where $a_1$ is a constant of integration. Using the third boundary condition in \Eref{eq:bcdFxib}, we find $a_1=0$.

From \Cref{eq:IEPM12} with $a_1=0$, \Cref{G} can be written as
\begin{equation}
\label{Gnew}
    G(\xi) = -\dfrac{1}{2} \xi F(\xi),
\end{equation}
and we see from \Eref{vprof} that $G$ describes the similarity profile for $v(x,y)$.
From \Cref{shearnew},
\begin{equation}
\label{shearnew1}
    -\dfrac{1}{\sqrt{Re_T}} \dfrac{G(\xi)}{x} =\bar{\tau}_{x y}+\bar{\tau}^T_{x y},
\end{equation}
and so $-G$ also gives the similarity profile for the shear stresses scaled by $1/\sqrt{Re_T}$.

To recover characteristic lengths, we introduce further scalings
\begin{equation}\label{eq:recoverlengths}
    F = A\bar{F}, \quad \xi = B\bar{\xi},  \quad G = A B \bar{G},
\end{equation}
so that \Eref{eq:IEPM12} and \Eref{eq:ConservedQuantityITOF} become
\begin{equation}\label{eq:ExtPrandtlPreTrans}
 \bar{\xi}\;\bar{F} + \frac{2 \beta}{B^2} \deriv{\bar{F}}{\bar{\xi}} +  \frac{2 \alpha A K_1^2}{B^3}\deriv{\bar{F}}{\bar{\xi}}\left[\left(\deriv{\bar{F}}{\bar{\xi}}\right)^2 + \frac{K_2^2}{B^2} \left(\deriv[2]{\bar{F}}{\bar{\xi}}\right)^2\right]^{1/2} = 0, 
\end{equation}
and
\begin{equation}\label{eq:ConservedQuantityITOFTrans}
 A B   \int_0^{{\bar{\xi_b}}} \bar{F} d\bar{\xi} = \dfrac{D}{2},
\end{equation}
respectively, where $\bar{\xi_b} = \xi_b/B$. Choosing 
\begin{equation}\label{eq:scalingschoice}
A = \frac{D^{3/4}}{2 \alpha^{1/4}K_1^{1/2}}, \qquad  B=D^{1/4} \alpha^{1/4} K_1^{1/2},
\end{equation}
and defining
\begin{equation}\label{eq:Transforms}
    \tilde{\beta} = \frac{2 \beta}{D^{1/2} \alpha^{1/2} K_1}, \quad \tilde{K}_2^2 =  \frac{K^2_2}{D^{1/2} \alpha^{1/2}K_1},  \quad 
\end{equation}
Equations~(\ref{eq:ExtPrandtlPreTrans}) and~(\ref{eq:ConservedQuantityITOFTrans}) simplify to 
\begin{equation}\label{eq:ExtPrandtlnunonzerofinal}
 \bar{\xi}\;\bar{F} + \tilde{\beta}\deriv{\bar{F}}{\bar{\xi}} +  \deriv{\bar{F}}{\bar{\xi}}\left[\left(\deriv{\bar{F}}{\bar{\xi}}\right)^2 + \tilde{K}_2^2 \left(\deriv[2]{\bar{F}}{\bar{\xi}}\right)^2\right]^{1/2} = 0,
\end{equation}
and
\begin{equation}\label{eq:ConservedQuantityFinal}
    \int_0^{\bar{\xi_b}} \bar{F} d\bar{\xi} = 1,
\end{equation}
respectively. The two remaining boundary conditions in \Eref{eq:bcdFxib} become
\begin{equation}\label{eqn:BCs_final}
    \bar{F}(\bar{\xi_b}) = 0, \ \ \deriv{\bar{F}}{\bar{\xi}}
    (\bar{\xi_b}) = 0.
\end{equation}

Note that Prandtl's original version of the extended model (i.e., where kinematic viscosity neglected) is recovered when $\tilde \beta=0$. Whether $\tilde \beta$ is zero or not, we see from the definitions in~\Eref{eq:Transforms} that the wake profile depends only on the product $DK_1^2$, and not on the conserved quantity $D$ and the proportionality constant $K_1$ independently.  Other choices of $A$ and $B$ do not seem to reveal any other interesting length scales. 

Finally, with the scalings~(\ref{eq:scalingschoice}), \Cref{Gnew} becomes
\begin{equation}
\label{Gnew1}
    \bar{G} (\bar{\xi}) = -\dfrac{1}{2} \bar{\xi} \bar{F}(\bar{\xi}),
\end{equation}
\revone{and from~(\ref{shearnew1}), the scaled similarity profile for the shear stress, which we denote by $\bar{g}(\bar{\xi})$, is
\begin{equation}
\label{shearnew11}
\bar{g}(\bar{\xi}) =    \dfrac{1}{2\sqrt{Re_T}} \bar{\xi} \bar{F}(\bar{\xi}).
\end{equation}}
\section{Results}\label{sec5}
\subsection{Similarity solutions}

\begin{figure}[t]
\hspace*{-5pt}\includegraphics[height=142pt,trim={.5cm .5cm .5cm 1cm},clip]{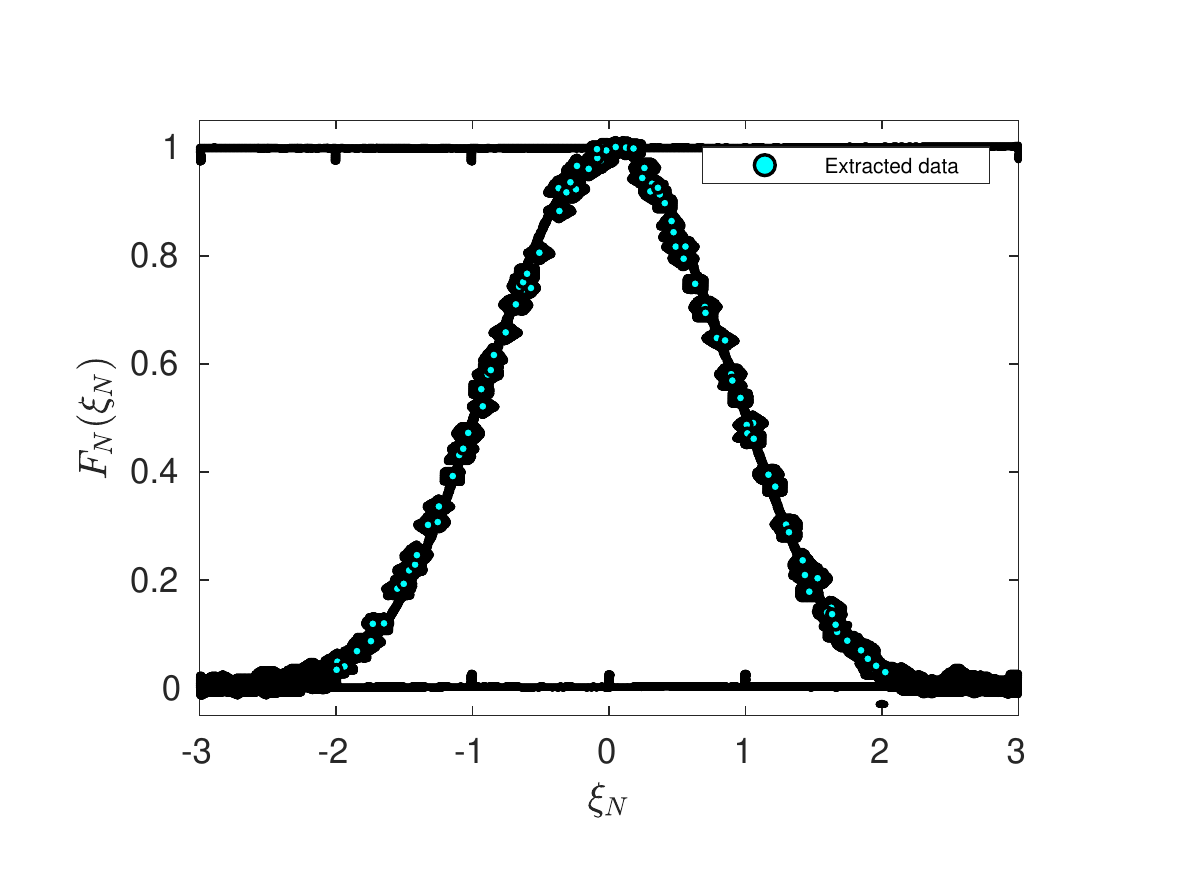}\hspace*{-18pt}  \begin{picture}(400,153)
\put(0,0){\includegraphics[height=142pt,trim={.5cm .5cm .5cm 1cm},clip]{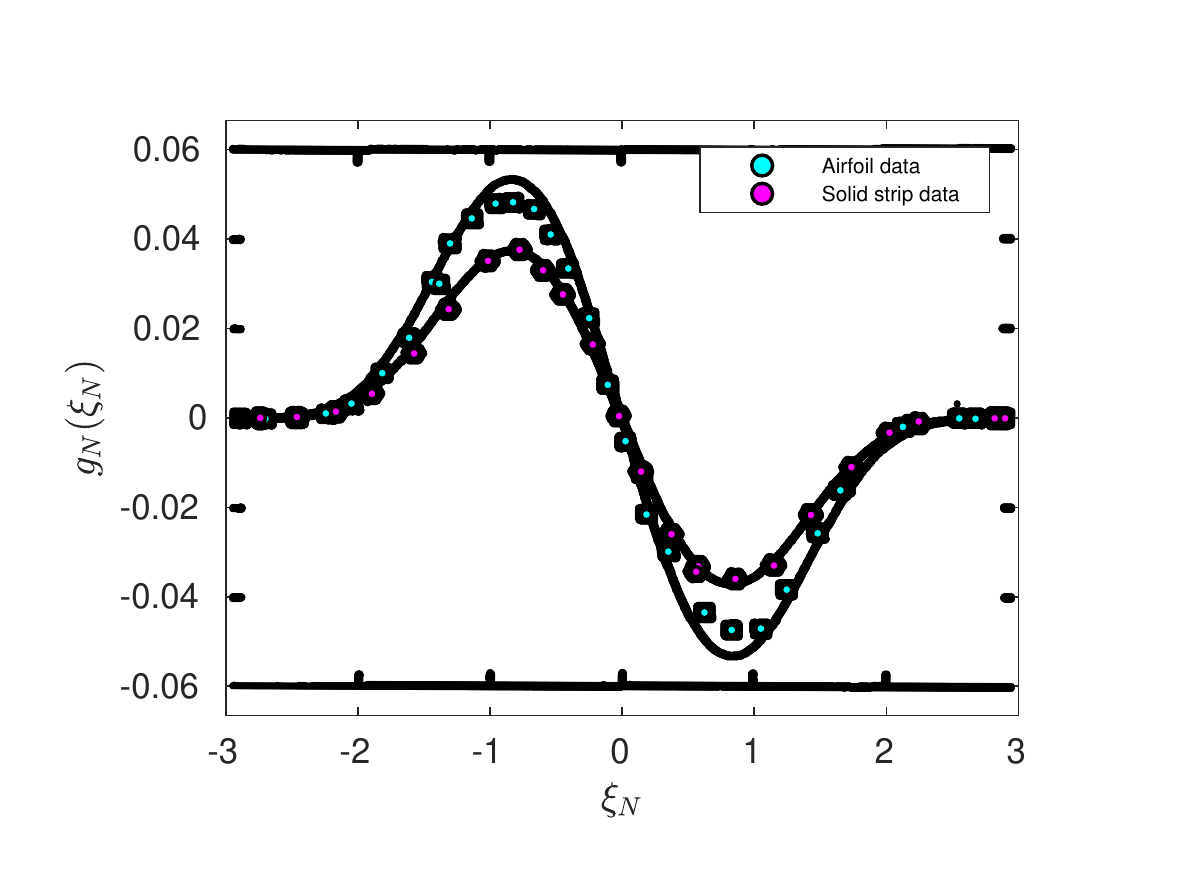}}
\put(55,125){\tiny Airfoil}
\put(70,126.5){\line(3,-1){10}}
\put(78,84){\tiny Solid strip}
\put(87,89){\line(-3,2){12}}
\end{picture}
\caption{Left: Experimental data and fitted curve~(\ref{eq:DataCurve}) for the normalised mean velocity profile.  Different icons correspond to different wake generators, which collapse to a single curve after appropriate normalisation. Equation~(\ref{eq:DataCurve}) provides a reasonable fit to the data, but is a heuristic approximation, rather than the result of a mathematical model. The raw data used to produce the figure is no longer available and the coloured dots represent digitally extracted values used in our analysis below. Values were only extracted when the corresponding data point was clearly identifiable. \revone{Right: Shear stress measurements for two different generators. The universality in the mean velocity profile is not present here, however the shear stress can be related to the normalised mean velocity by~(\ref{datastress}).} (Both figures are reproduced with permission of the corresponding author of~\cite{Wygnanski}.)}
\label{fig:Wygnanskfigure3}
\end{figure}
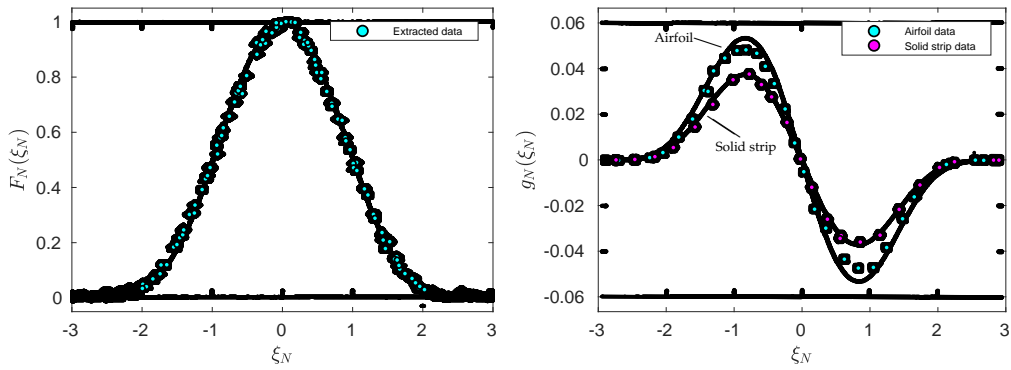

From data generated from wind tunnel experiments, Wygnanski et al.~\cite{Wygnanski} demonstrated that regardless of the wake generator --- examples of which included cylinders, symmetric airfoils, and flat plates ---  the shape of the {\em normalised} mean velocity profile far downstream is the same. 
In particular, they show that the curve
\begin{equation}
    F_N(\xi_N) = \exp{\left[-0.637\xi_N^2-0.056\xi_N^4\right]}, \label{eq:DataCurve}
\end{equation}
provides a good fit to the mean velocity profile of the far wake when the similarity variables are normalised so that $F_N(0)=1$ and $F_N(1)=1/2$.
\revone{This universality is not true of the shear stresses, (\ref{shearnew11}), whose profiles depend on the wake generator.}
\revone{From \cite{Wygnanski}, the normalised Reynolds shear stress $g_N$ is given by
\begin{equation}
\label{datastress}
    g_N (\xi_N) = S \xi_N F_N(\xi_N),
\end{equation}
where the constant $S$ depends on the wake generator (compare with~\Cref{shearnew11}).} The data and the curves~(\ref{eq:DataCurve}) \revone{and (\ref{datastress})} are shown in Figure~\ref{fig:Wygnanskfigure3}.

We now compare the various closure models presented previously with this experimental data. For each model, we solve for the normalised mean velocity profile, $F_N(\xi_N)$, \revone{and use \Cref{datastress} to determine the shear stresses. The airfoil and solid strip in Figure~\ref{fig:Wygnanskfigure3} correspond to $S = 0.103$ and $S = 0.072$, respectively.}
Unfortunately, the raw data used to produce the original version of  Figure~\ref{fig:Wygnanskfigure3} is no longer available, so we manually extract values from the figures in~\cite{Wygnanski} to allow a quantitative comparison. The extracted values used have been made available online~\cite{HALEEPML}. 

\subsection{Solving the models}\label{subsec:solving}
The CEV model, which results in the boundary value problem~\cite{P}
\begin{equation}
 \deriv[2]{F}{\xi} +   \deriv{}{\xi} \left[\xi F(\xi) \right]  = 0, \quad 
    F(\infty) = 0, \ \ \ F'(\infty) = 0,
\end{equation}
has a closed-form solution, and the resulting normalised mean velocity profile is given by~\cite{Wygnanski}
\begin{equation}
    F_N(\xi_N)=\exp\left[-\ln(2) \xi_N^2\right].\label{eqn:CEVfit}
\end{equation}

An analytic solution of~\Eref{eq:ExtPrandtlnunonzerofinal}--\Eref{eqn:BCs_final} can be obtained for the case $\tilde \beta=\tilde K_2=0$ (i.e. PML with $\tilde \beta = 0$), giving the normalised profile~\cite{Cafiero}
\begin{equation}
    F_N(\xi_N) =  \left[\left(\dfrac{\xi_N}{\xi_b} \right)^{3/2} - 1 \right]^2, \ \ \xi_b  = \left(2 \left(3 + 2 \sqrt{2}\right)\right)^{1/3}.
\end{equation}
Notice here that the condition $F_N(\xi_b)=0$ results in a finite wake boundary, and hence the wake is bounded in the $y$-direction.
This is unlike in the CEV model where the exponential solution does not vanish for finite $\xi_b$.

\begin{figure}[t!]\centering
\includegraphics[width=\textwidth]{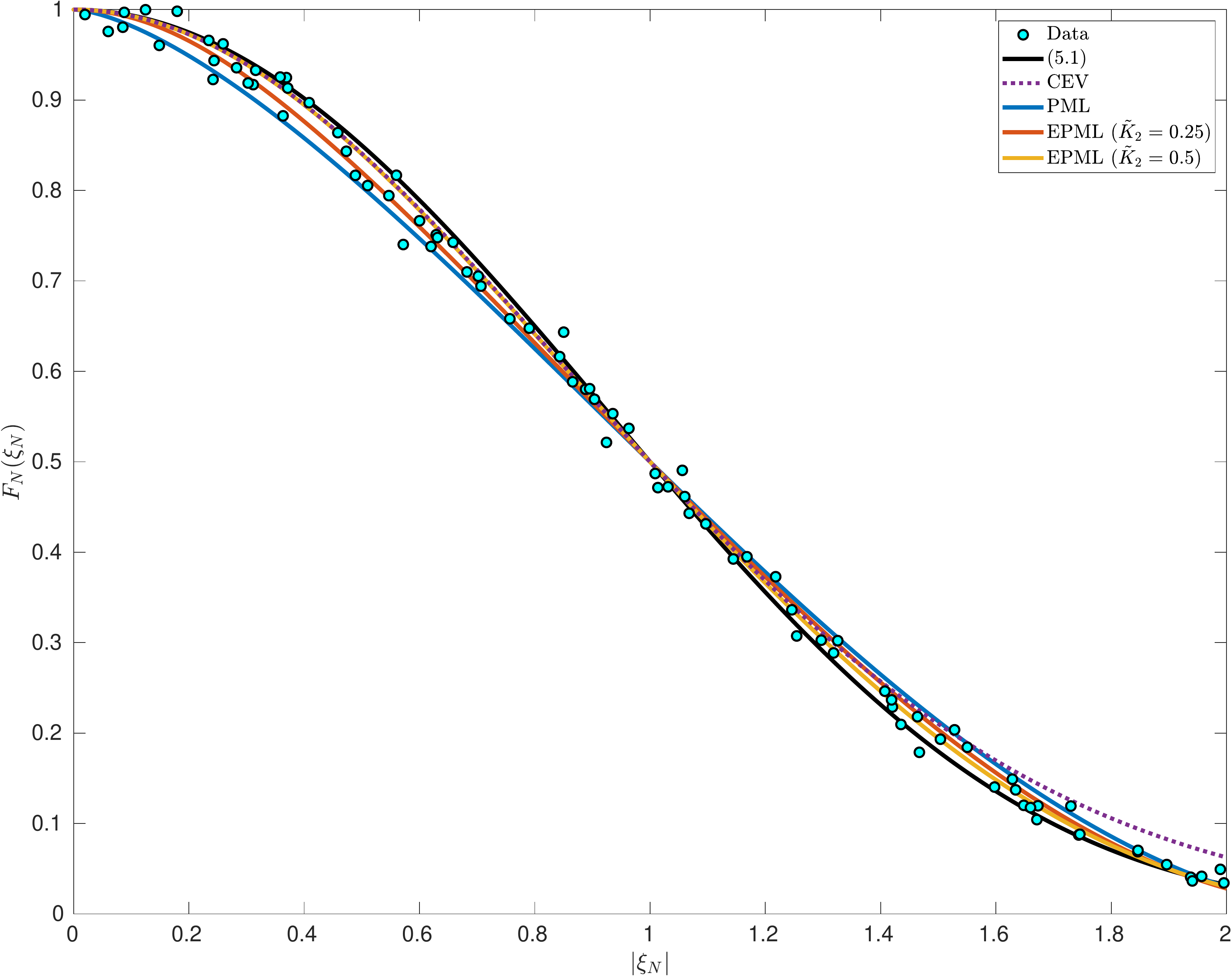}
\caption{Normalised mean velocity profiles of various closure models compared to experimental data from~\cite{Wygnanski}. The CEV and PML ($\tilde \beta = 0.01$) models provide a poor fit to the experimental data, particularly near the tail of the wake. As $\tilde K_2$ is increased from zero, the EPML ($\tilde \beta = 0.01$) solutions increase for $\xi_N < 1$ and decrease for $\xi_N > 1$, more closely matching the heuristic fit~(\ref{eq:DataCurve}) and the data. It is interesting to note that the EPML solution with $\tilde K_2 = 0.5$ matches almost exactly the solution from the CEV model for $0 \le \xi_N \le 1$. The best fit to the data is obtained when $\tilde K_2$ is around 0.375  (see Table~\ref{tab:l2normtable}).}\label{figaa}
\end{figure}

Unfortunately, other than this special case, a closed form solution of~\Eref{eq:ExtPrandtlnunonzerofinal}--\Eref{eqn:BCs_final} seems beyond reach and the ODE must be solved numerically. For both the PML model with $\tilde \beta > 0$ and for the EPML model, we discretise the equations using a Hermite pseudospectral method~\cite{DMSUITE}.\footnote{Although a Laguerre spectral method would be more typical for a semi-infinite domain, we find the Hermite to be more accurate and reliable in solving~\Eref{eq:ExtPrandtlnunonzerofinal}--\Eref{eqn:BCs_final}. We suspect this is due to the super-exponential decay of the solution (more closely matching the Hermite weight) and because the Laguerre method requires enforcing an additional boundary condition at the origin. MATLAB code to solve~\Eref{eq:ExtPrandtlnunonzerofinal}--\Eref{eqn:BCs_final} and reproduce Figure~\ref{figaa} is available in an online repository~\cite{HALEEPML}.}  The resulting nonlinear system is then  solved using a Newton iteration with initial guess~(\ref{eq:DataCurve}).  
The results are shown in Figure~\ref{figaa}, along with the data and fit from~\cite{Wygnanski}, and the CEV solution~(\ref{eqn:CEVfit}).  
For both PML and EPML we find that increasing $\tilde \beta$ increases the effective width of the wake, but not significantly, and so fix $\tilde \beta = 0.01$.

We see, as expected, that the CEV model provides a good fit near the centre line of the wake, but overestimates near the wake boundary. The PML model also overestimates near the wake boundary, and underestimates near the centre line of the wake. As $\tilde K_2$ is increased from 0 the EPML solution increases near the centre line, and decreases near the wake boundary, hence providing a significantly improved match to both the curve~(\ref{eq:DataCurve}) and the experimental results from~\cite{Wygnanski}. Increasing $\tilde K_2$ beyond 0.5 might give an even closer match to~(\ref{eq:DataCurve}), but the numerical solution begins to break down for $\tilde K_2>0.5$, we suspect due to a lack of regularity. Since~(\ref{eq:DataCurve}) is only a heuristic approximation, and values of $0.25 < \tilde K_2 <  0.4$ give a better fit to the data itself (see Table~\ref{tab:l2normtable}), we do not pursue this further. However, it is interesting to note that for $0 \le \xi_N \le 1$ the EPML model solution with $\tilde K_2 = 0.5$ corresponds almost exactly with the CEV model solution.

\begin{table}[t]
    \centering
    \begin{tabular}{c|cccccccccccccc}
Model: & (\ref{eq:DataCurve}) & CEV & PML & $\tilde K_2 = 0.1$ & $\tilde K_2 = 0.2$ & $\tilde K_2 = 0.3$ & $\tilde K_2$ = 0.4  &  $\tilde K_2 = 0.5$\\
$\ell_2$ error: & 0.181 & 0.194 & 0.206 &	 0.186 & 0.160 & 0.144 & 0.141 &	 0.148
\end{tabular}
    \caption{The 2-norm errors of the various closure models (CEV, PML, and EPML with $\tilde K_2 = 0.1, 0.2, \ldots, 0.5$) applied to the experimental data from~\cite{Wygnanski}. For values of $\tilde K_2$ in the range $[0.25, 5]$ the EPML model provides a significantly better fit (up to 27.3\%) than both the CEV and PML models, and is even an improvement (22.1\%) on the heuristic fit~(\ref{eq:DataCurve}). For completeness, we report that the smallest error was obtained with $\tilde K_2$ around 0.375.}
    \label{tab:l2normtable}
\end{table}

The success of the EPML model here may be attributed to two factors:  First, from the asymptotic solution as $y \rightarrow y_b(x)$ derived in Section 3, the second derivative, $F_N''$, tends to zero as $\xi \rightarrow \infty$, which is supported by the numerical results. Hence the eddy viscosity vanishes at the boundary of the wake, capturing the physical behaviour correctly. Second, there are two free parameters that can be chosen to fit the data. Although the PML model has the free parameter $\tilde \beta$, we have seen this has a minimal affect on the shape of the normalised mean velocity profile. The new parameter, $\tilde K_2$, corresponding to the second scaled mixing length in the EPML model, has a far more significant impact on the shape of the normalised profile.  Because the EPML model was derived using physical considerations and provides a significantly improved fit to the experimental data, we conclude that it outperforms the other models considered in this work \rev{when predicting the mean velocity profile}.

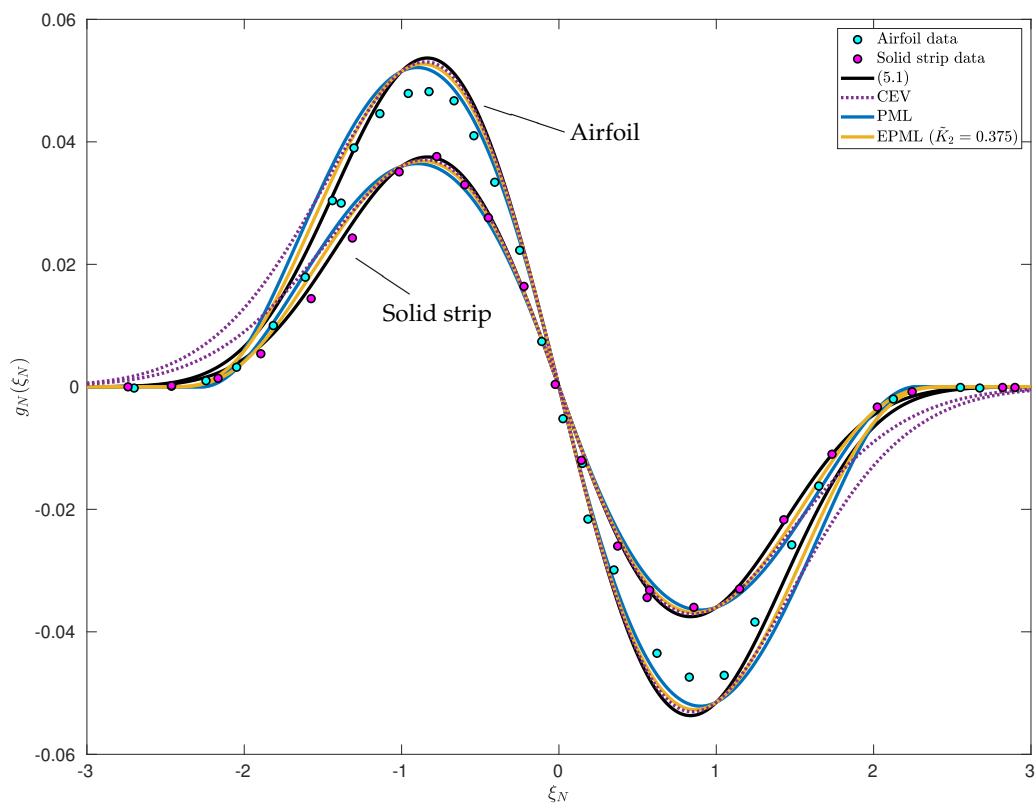
\begin{figure}[ht!]\centering
\begin{picture}(800,300)
\put(0,0){\includegraphics[width=\textwidth]{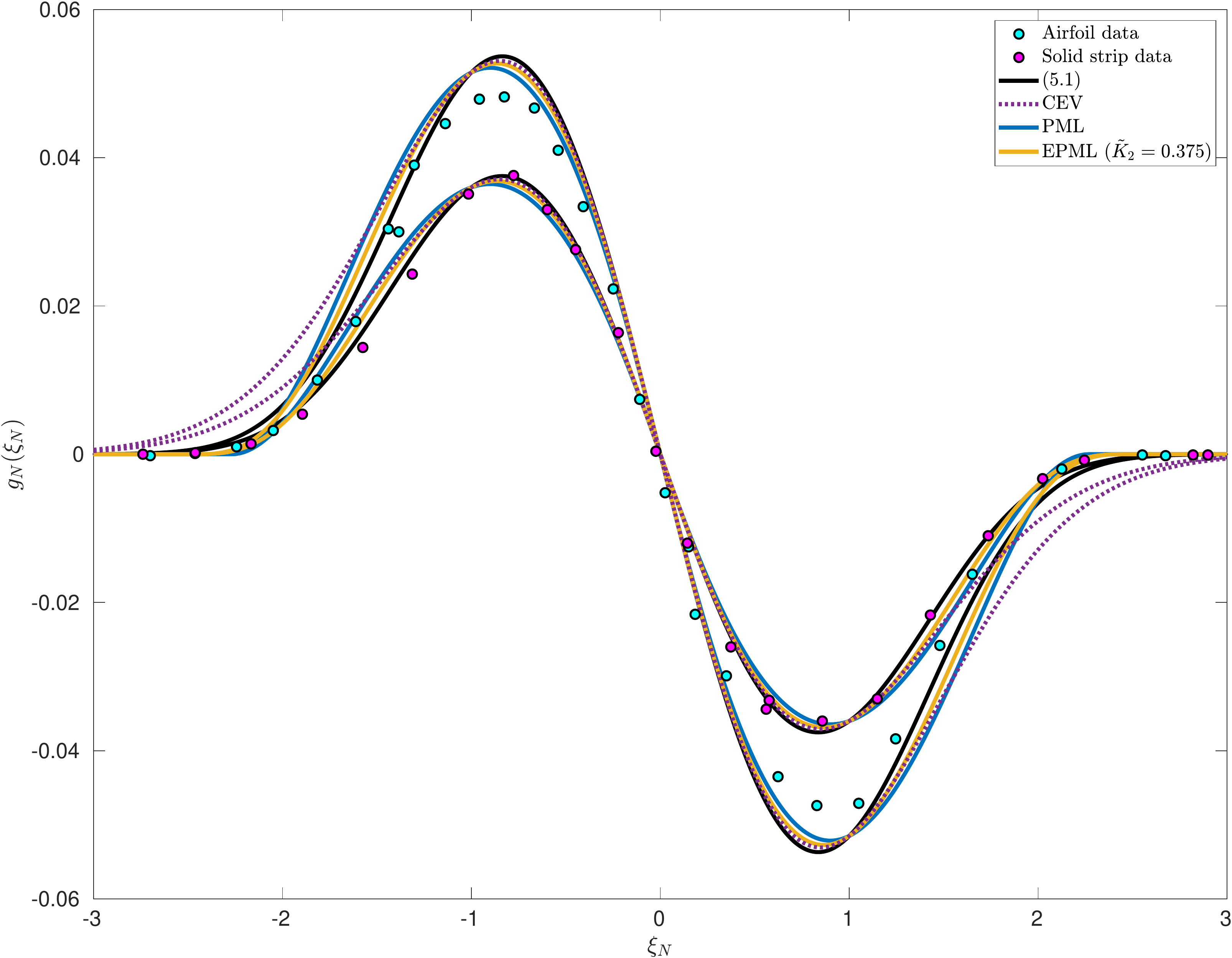}}
\put(210,250){Airfoil}
\put(208,253){\line(-3,1){30}}
\put(140,182){Solid strip}
\put(150,192){\line(-3,2){20}}
\end{picture}
\caption{\revone{Normalised shear stress profiles for the different closure models compared to experimental data from~\cite{Wygnanski}. For each closure model, the shear stresses are computed from the normalised mean velocity profiles using~(\ref{datastress}), with $S$ = 0.103 for the airfoil and $S$ = 0.072 for the strip. Similarly to the mean velocity profile, the CEV model significantly overestimates the magnitude of the stress towards the wake boundaries. Equation~(\ref{eq:DataCurve}) gives a good fit for the solid strip generator, but overestimates the maximum magnitude of the stress for the airfoil. The PML and EPML models also overestimate the stress for the airfoil.}}\label{fig5}
\end{figure}

\revone{In Figure~\ref{fig5} we perform a similar comparison, but now for the shear stresses. In particular, for each of the closure models (and the fit~(\ref{eq:DataCurve})) we substitute the obtained normalised mean velocity profiles into~(\ref{datastress}) and compare to the experimental data from Figure~\ref{fig:Wygnanskfigure3}. Whilst it is clear that the CEV model significantly overestimates the stresses at the wake boundaries, other conclusions are harder to draw. For the solid strip generator, all but the CEV give a reasonable fit to the data, with the fit~(\ref{eq:DataCurve}) being slightly better than the rest; particularly towards the tails of the wake. For the airfoil generator, all of the curves significantly overestimate the maximum magnitude of the stress. The fit~(\ref{eq:DataCurve}) is still the best towards the wake boundaries, but overestimates more than the other models near the centre line. Of the curves corresponding to mathematical models, the EPML (here with $K_2 = 0.375)$ improves on that of the PML (and CEV) models, but the improvement is not so pronounced as for the velocity profiles in Figure~\ref{figaa}. 
The fact that the models gave a good fit for the normalized velocity profiles but not the Reynolds shear stresses suggests that further investigation into the relationship~(\ref{datastress}) is required. Although we do not pursue this here, it raises an important question as to whether eddy viscosity models will ever be capable of accurately predicting the shear stresses.}

\section{Summary}\label{sec6}

The eddy viscosity closure model was used to complete the system of equations for the mean flow variables describing the far downstream two-dimensional turbulent classical wake, and the governing equations were expressed in terms of the mean velocity deficit in the $x$-direction.  The boundary conditions on the mean velocity deficit were obtained by imposing matching conditions between the turbulent wake region with the inviscid laminar mainstream flow, and the condition that the velocity deficit is maximised on the axis of the wake. Since the boundary conditions were independent of the choice of closure model, the conserved quantity was also independent. The $y$-component of the mean velocity was then derived and it was shown that there was no entrainment at the far wake boundary for finite-valued eddy viscosities. This depends critically on the linear approximation made for the inertia in \Cref{eq:finalequation}.

An outline of the derivation of Prandtl's mixing length model was provided, which was then modified to introduce a \revtwo{new} derivation of the extended Prandtl mixing length model.  Scaling solutions admitted by the governing equations of the extended model were obtained, and it was shown when kinematic viscosity is neglected that Prandtl's hypothesis is necessary to obtain a similarity solution. Conversely, when kinematic viscosity is included, no additional constraints are required. The similarity variables were used to reduce the PDE to an ODE, and exact solutions were found for special cases. Numerical results were calculated for the remaining cases, and the normalised similarity mean velocity \revone{and shear stress} profiles were compared for various closure models. \revone{For the mean velocity profiles,} it was shown that the extended Prandtl mixing length model gives a significantly improved fit to experimental data when compared with the other considered models. \revone{The shear stress profiles, which are computed from the mean velocity profiles, did not provide such a good fit to the experimental data, suggesting that more work is needed in relating the two.}

The work presented here also sets the stage for the development of new closure models \revone{and application of similarity methods in turbulence modelling}. Unlike the CEV model, the PML and EPML models both satisfy  $\nu_T \rightarrow 0$ as $y \rightarrow \pm y_b(x)$, which adheres to the condition of mainstream matching between the  laminar mainstream flow region and the turbulent wake.  We saw that for models satisfying this property, the solution tends to the exponential solution for the laminar wake. The exponential solution satisfies  $\partial^n \bar{w}/\partial y^n \rightarrow 0$ as $y \rightarrow \pm \infty$ for $n \in \mathbb{Z}^+$, meaning that an entire class of closure models {with nonzero kinematic viscosity and} depending on partial derivatives higher than  $\partial^2 \bar{w}/\partial y^2$ that satisfy the mainstream matching condition might be produced. This is an interesting consideration for future work.

\enlargethispage{20pt}



 \aucontribute{AJH lead the development of the work and was responsible for many of the ideas presented in Sections 2, 3. NH produced the numerical solutions and analysis, and with AJH, lead the structuring and writing of the paper. KB produced all diagrams in Sections 2 and 3, obtained the scaling solutions in Section 4, and provided valuable input into the other sections. DPM, as the primary subject matter expert, made many corrections to all sections, significantly improving upon the quality of the work.
 }


\funding{DPM thanks the National Research Foundation, Pretoria, South Africa, for financial support. Grant number: 96270}

\ack{The authors are grateful to I.\ Wygnanski (University of Arizona) for permitting the reproduction of Figure~\ref{fig:Wygnanskfigure3} and to the anonymous referees for their useful feedback. We also thank P. Broadbridge (La Trobe University) for his valuable comments.}


\bibliographystyle{RS.bst} 
\bibliography{main.bib}

\begin{thebibliography}{99}

\bibitem{Ten}
Tennekes H, Lumley JL. 1971 {\em A First Course in Turbulence}.
Cambridge, Massachusetts, and London: MIT Press.

\bibitem{P}
Pope SB. 2000 {\em Turbulent Flows}.
Cambridge University Press.

\bibitem{Turbine}
Howland MF, Lele SK, Dabiri JO. 2019  Wind farm power optimization through wake
  steering. {\em Proceedings of the National Academy of Sciences} \textbf{116},
  14495--14500.

\bibitem{Breton}
Breton SP, Sumner J, Sørensen JN, Hansen KS, Sarmast S, Ivanell S. 2017  A
  survey of modelling methods for high-fidelity wind farm simulations using
  large eddy simulation. {\em Philosophical Transactions of the Royal Society
  A: Mathematical, Physical and Engineering Sciences} \textbf{375}, 20160097.

\bibitem{Mehta}
Mehta D, van Zuijlen AH, Koren B, Holierhoek JG, Bijl H. 2014  Large Eddy
  Simulation of wind farm aerodynamics: A review. {\em Journal of Wind
  Engineering and Industrial Aerodynamics} \textbf{133}, 1--17.

\bibitem{winda}
Jim{\'e}nez {\'A}, Crespo A, Migoya E. 2010  Application of a LES technique to
  characterize the wake deflection of a wind turbine in yaw. {\em Wind Energy}
  \textbf{13}, 559--572.

\bibitem{windb}
Abkar M, Dabiri JO. 2017  Self-similarity and flow characteristics of
  vertical-axis wind turbine wakes: an LES study. {\em Journal of Turbulence}
  \textbf{18}, 373--389.

\bibitem{windc}
Wu YT, Porté-Agel F. 2015  Modeling turbine wakes and power losses within a
  wind farm using LES: An application to the Horns Rev offshore wind farm. {\em
  Renewable Energy} \textbf{75}, 945 -- 955.

\bibitem{windd}
Göçmen T, van~der Laan P, Réthoré PE, Diaz AP, Larsen GC, Ott S. 2016  Wind
  turbine wake models developed at the technical university of Denmark: A
  review. {\em Renewable and Sustainable Energy Reviews} \textbf{60}, 752--769.

\bibitem{Hutter2016}
Hutter K, Wang Y. 2016 {\em Turbulent Mixing Length Models and Their
  Applications to Elementary Flow Configurations}.
Springer International Publishing.

\bibitem{Cafiero}
Cafiero G, Obligado M, Vassilicos JC. 2020  Length scales in turbulent free
  shear flows. {\em Journal of Turbulence} \textbf{21}, 243--257.

\bibitem{Note}
Doshi MR, Gill WN. 1970  A note on the mixing length theory of turbulent flow.
  {\em AIChE Journal} \textbf{16}, 885--888.

\bibitem{Reynolds}
Reynolds O. 1895  IV. On the dynamical theory of incompressible viscous fluids
  and the determination of the criterion. {\em Philosophical Transactions of
  the Royal Society of London. (A.)} \textbf{186}, 123--164.

\bibitem{Liepmann}
Liepmann HW. 1952  {Aspects of the turbulence problem}. {\em Zeitschrift
  f{\"u}r angewandte Mathematik und Physik} \textbf{3}, 407--426.

\bibitem{Boussinesq1}
Boussinesq J. 1877  Essai sur la th\'eorie des eaux courantes. {\em M\'emoires
  pr\'esent\'es par divers savants \`a l'Acad\'emie des Sciences}
  \textbf{XXIII}.

\bibitem{Goldstein1933}
Goldstein S. 1933  On the two-dimensional steady flow of a viscous fluid behind
  a solid body. {\em Proceedings of the Royal Society of London. Series A,
  Containing Papers of a Mathematical and Physical Character} \textbf{142},
  545--562.

\bibitem{Wygnanski}
Wygnanski I, Champagne F, Marasli B. 1986  On the large-scale structures in
  two-dimensional, small-deficit, turbulent wakes. {\em Journal of Fluid
  Mechanics} \textbf{168}, 31--71.

\bibitem{Prandtl}
Prandtl L. 1925  {Bericht {\"u}ber Untersuchenden zur ausgebildeten Turbulenz}.
  {\em Zeitschrift f{\"u}r angewandte Mathematik und Mechanik} \textbf{5},
  136--139.
English: NACA-TM-1231.

\bibitem{Swain}
Swain LM. 1929  On the turbulent wake behind a body of revolution. {\em
  Proceedings of the Royal Society of London. Series A, Containing Papers of a
  Mathematical and Physical Character} \textbf{125}, 647--659.

\bibitem{PrandtlHypothesis}
Prandtl L. 1927  {{\"U}ber die ausgebildete Turbulenz}. In {\em Verhandlungen
  des II. Internationalen Kongresses f{\"u}r Technische Mechanik 1926} pp.
  62--75. Z{\"u}rich:F{\"u}{\ss}li-Verlag.

\bibitem{Luo}
Luo X, Liu Pl, Luo Ha. 2008  Improvement of Prandtl mixing length theory and
  application in modeling of turbulent flow in circular tubes. {\em Journal of
  Central South University of Technology} \textbf{15}, 774--778.

\bibitem{ashex}
Hutchinson AJ. 2019-20  The extended Prandtl closure model applied to the
  two-dimensional turbulent classical far wake. {\em MATRIX Annals, MATRIX Book
  Series, Springer} \textbf{4}.
To appear.

\bibitem{ash3}
Hutchinson AJ, Mason DP. 2015  Revised Prandtl mixing length model applied to
  the two-dimensional turbulent classical wake. {\em International Journal of
  Non-Linear Mechanics} \textbf{77}, 162--171.

\bibitem{cantwell_1978}
Cantwell BJ. 1978  Similarity transformations for the two-dimensional,
  unsteady, stream-function equation. {\em Journal of Fluid Mechanics}
  \textbf{85}, 257–271.

\bibitem{oberlack_2001}
Oberlack M. 2001  A unified approach for symmetries in plane parallel turbulent
  shear flows. {\em Journal of Fluid Mechanics} \textbf{427}, 299–328.

\bibitem{oberlack_2007}
Razafindralandy D, Hamdouni A, Oberlack M. 2007  New turbulence models
  preserving symmetries. In {\em 5th International Symposium on Turbulence and
  Shear Flow Phenomena}.

\bibitem{oberlack_2015}
Oberlack M, Wac\l{}awczyk M, Rosteck A, Avsarkisov V. 2015  Symmetries and
  their importance for statistical turbulence theory. {\em Mechanical
  Engineering Reviews} \textbf{2}, 15--157.

\bibitem{Ibragimov}
Ibragimov NH, {\"U}nal G. 1994  Lie groups in turbulence: I. Kolmogorov’s
  invariant and the algebra $l_{\tau}$. {\em Lie Groups and their Applications}
  \textbf{1}, 98–103.

\bibitem{Unal}
{\"U}nal G. 1994  Application of equivalence transformations to inertial
  subrange of turbulence. {\em Lie Groups and their Applications} \textbf{1},
  232--240.

\bibitem{schlichtingGersten}
Schlichting H, Gersten K. 2017 {\em Boundary-Layer Theory}.
Springer Berlin Heidelberg ninth edition.

\bibitem{Schlichting}
Schlichting H. 1979 {\em Boundary-Layer Theory}.
Springer Berlin Heidelberg seventh edition.

\bibitem{ash}
Hutchinson AJ, Mason DP, Mahomed FM. 2015  Solutions for the turbulent
  classical wake using Lie symmetry methods. {\em Communications in Nonlinear
  Science and Numerical Simulation} \textbf{23}, 51--70.

\bibitem{AshVorticity}
Hutchinson AJ. 2018  Application of a modified Prandtl mixing length model to
  the turbulent far wake with a variable mainstream flow. {\em Physics of
  Fluids} \textbf{30}, 095102.

\bibitem{ash2}
Hutchinson AJ, Mason DP. 2016  Lie symmetry methods applied to the turbulent
  wake of a symmetric self-propelled body. {\em Applied Mathematical Modelling}
  \textbf{40}, 3062--3080.

\bibitem{ash4}
Hutchinson AJ. 2016  A unified theory for turbulent wake flows described by
  eddy viscosity. {\em International Journal of Non-Linear Mechanics}
  \textbf{81}, 40 -- 54.

\bibitem{Brad}
Bradshaw P. 1974  Possible origin of Prandt's mixing-length theory. {\em
  Nature} \textbf{249}, 135--136.

\bibitem{Dres}
Dresner L. 1983 {\em Similarity solutions of nonlinear partial differential
  equations}.
Pitman, Boston.

\bibitem{S}
Sj{\"o}berg A. 2007  Double reduction of PDEs from the association of
  symmetries with conservation laws with applications. {\em Applied Mathematics
  and Computation} \textbf{184}, 608--616.

\bibitem{HALEEPML}
Hale N. 2020  MATLAB code for the EPML model.
  \url{https://github.com/nickhale/EPML}.

\bibitem{DMSUITE}
Weideman J, Reddy S. 2000  A MATLAB Differentiation Matrix Suite. {\em ACM
  Trans. Math. Softw.} \textbf{26}, 465–519.

\end{thebibliography}
\end{document}